\documentclass[journal,10pt]{IEEEtran}

\usepackage{amsthm}
\usepackage{amsfonts}
\usepackage{amssymb}
\usepackage{mathrsfs}
\usepackage{mathtools}
\usepackage{float}
\usepackage[pdftex]{graphicx}
\usepackage{amsmath}
\usepackage{color}
\usepackage{multirow}
\usepackage{graphicx}
\usepackage{graphicx,float,wrapfig,epstopdf,amsmath}
\usepackage{multirow}

\usepackage{comment}
\usepackage{balance}
\usepackage{lipsum}
\usepackage[inline]{enumitem}
\usepackage{cuted}
\usepackage[caption=false,font=footnotesize]{subfig}
\usepackage{cite}
\usepackage{hyperref}
\usepackage{enumitem}
\usepackage[]{algorithmicx}
\usepackage{algpseudocode,algorithm}

\usepackage[T1]{fontenc}

\newlist{myenumi}{description}{10}
\setlist[myenumi]{labelindent=\parindent, leftmargin=*, label=(\roman*), align=left}
\setlist[myenumi]{leftmargin=0pt}
\newtheorem{theorem}{Theorem}

\newtheorem{lemma}[theorem]{Lemma}

\hypersetup{%
	colorlinks=true,
	linkcolor={black},
	raiselinks=false
}
\DeclareMathOperator*{\maximize}{max} 
\DeclareMathOperator*{\minimize}{min} 
\DeclareMathOperator*{\subjectto}{s.\hspace{3pt} t.:\hspace{3pt}} 

\begin{document}
	
	
	
	\title{Terahertz-Band Joint Ultra-Massive MIMO Radar-Communications: Model-Based and Model-Free Hybrid Beamforming}
	
	\author{\IEEEauthorblockN{Ahmet M. Elbir, \textit{Senior Member IEEE}, Kumar Vijay Mishra, \textit{Senior Member IEEE}}, and Symeon Chatzinotas, \textit{Senior Member IEEE}
		
		\thanks{This work was supported in part by the  ERC Project  AGNOSTIC.}
		\thanks{A. M. E. is with the Department of Electrical and Electronics Engineering, Duzce University, Duzce, Turkey, and  SnT, University of Luxembourg,	Luxembourg City, Luxembourg (e-mail: ahmetmelbir@gmail.com).} 
		\thanks{K. V. M. is with the United States Army Research Laboratory, Adelphi, MD 20783 USA (e-mail: kumarvijay-mishra@uiowa.edu).}
		\thanks{S. C. is with the SnT at the University of Luxembourg, Luxembourg (e-mail: symeon.chatzinotas@uni.lu). }
	}

	\maketitle

	\begin{abstract}
		Wireless communications and sensing at terahertz (THz) band are increasingly investigated as promising short-range technologies because of the availability of high operational bandwidth at THz. In order to address the extremely high attenuation at THz, ultra-massive multiple-input multiple-output (MIMO) antenna systems have been proposed for THz communications to compensate propagation losses. However, the cost and power associated with fully digital beamformers of these huge antenna arrays are prohibitive.	In this paper, we develop  wideband hybrid beamformers based on both model-based and model-free techniques for a new group-of-subarrays (GoSA) ultra-massive MIMO structure in low-THz band. Further, driven by the recent developments to save the spectrum, we propose beamformers for a joint ultra-massive MIMO radar-communications system, wherein the base station serves multi-antenna user equipment (RX), and tracks radar targets by generating multiple beams toward both RX and the targets. We formulate the GoSA beamformer design as an optimization problem to provide a trade-off between the unconstrained communications beamformers and the desired radar beamformers. To mitigate the beam split effect at THz band arising from frequency-independent analog beamformers, we propose a phase correction technique to align the beams of multiple subcarriers toward a single physical direction. Additionally, our design also exploits second-order channel statistics so that an infrequent channel feedback from the RX is achieved with less channel overhead. To further decrease the ultra-massive MIMO computational complexity and enhance robustness, we also implement deep learning solutions to the proposed model-based hybrid beamformers. Numerical experiments demonstrate that both techniques outperform the conventional approaches in terms of spectral efficiency and radar beampatterns, as well as exhibiting less hardware cost and computation time.
	\end{abstract}
	
	
	\section{Introduction}
	\label{sec:intro}
	\begin{figure*}[t]
		\centering
		{\includegraphics[draft=false,width=\textwidth]{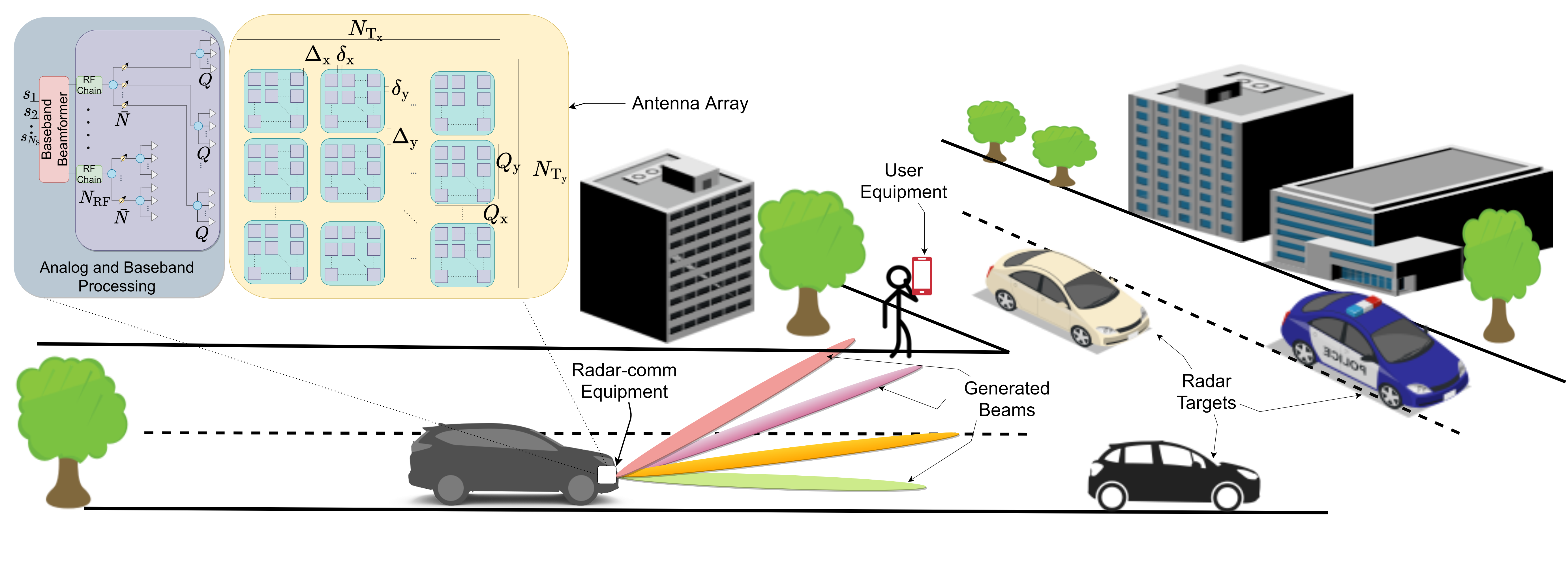} }
		
		\caption{A radar-communications system for a vehicle-to-vehicle (V2V) and vehicle-to-device (V2D) scenario, wherein a single THz radar-communications unit, with a $N_\mathrm{T} = N_{\mathrm{T}_\mathrm{x}}\times  N_{\mathrm{T}_\mathrm{y}}$ antenna array, is mounted onto a vehicle to simultaneously transmit toward both communications receiver and vehicular targets.}
		\label{figHB_TX}
	\end{figure*}
	
	In recent years, millimeter-wave (mmWave) spectrum has been extensively  studied to address the demands for high data rates in the fifth-generation (5G) wireless communications \cite{heath2016overview,mishra2019toward}. The maximum mmWave bandwidth being tens of GHz, it is not possible to achieve data rates of the order of terabits-per-second (Tb/s) without significantly enhancing the current physical-layer efficiency. As a result, the future sixth-generation (6G) networks are expected to exploit the THz frequencies ($0.3$-$10$ THz) \cite{rajatheva2020white,song2011present}, where hundreds of GHz bandwidth is available to facilitate Tb/s rates without dramatic efficiency increase in the physical-layer \cite{akyildiz2016realizing}. There is, therefore, considerable interest in developing THz wireless solutions \cite{kurner2014towards,tekbiyik2019terahertz}.
	
	Higher bandwidths also result in improved radar range resolution. At present, mmWave radars with a few GHz bandwidths such as those used in automotive applications \cite{dokhanchi2019mmwave} at $24$ and $77$ GHz are unable to yield high-resolution images compared to the optical sensors \cite{wang2020displaced}. Higher operating frequency have smaller antenna apertures and microwave components, which is beneficial for radar deployment on cost and area-sensitive platforms such as drones and ground vehicles. At THz, the physical aperture of the antenna is expected to be very small and the availability of large transmit bandwidth has the potential to offer image resolutions closer to that from the optical sensors \cite{marchetti2018radar}.
	
	The precise definition of THz band varies among different community members. Recent works in wireless communications generally define this band in the range $0.03$-$10$ THz \cite{faisal2020ultramassive} with an obvious overlap with the conventional mmWave frequencies. For the radar, microwave, and remote sensing engineers, THz band starts at the \textcolor{black}{upper-mmWave limit of $100$ GHz and, in particular, \textit{low-THz} term is used} for the range $0.1$-$1$ THz \cite{norouzian2017low}. In optics, on the other hand, THz spectrum is defined to end at $10$ THz, beyond which frequencies are considered far-infrared \cite{schmuttenmaer2004exploring}. The Terahertz Technology and Applications Committee of the IEEE Microwave Theory and Techniques Society (MTT-S) focuses on $0.3$-$3$ THz range while the IEEE Transactions on Terahertz Science and Technology journal targets $0.3$-$10$ THz. In this paper, in order to be consistent with the developments in THz radar and communications, our work is relevant largely to the low-THz frequencies used by the remote sensing community.
	
	The low-THz spectrum lies in the \textit{atmospheric window} - a region of local minimum  attenuation \cite{sun2016predicting,piesiewicz2007scattering}. However, the associated path losses are so high that low-THz radar applications have conventionally been limited to indoor environments such as vital sign monitoring \cite{xu2014vital}, food contamination detection \cite{ok2014high}, and airport security \cite{appleby2007standoff}. Recent characterizations of low-THz radars carried out in the outdoor environments for automotive applications \cite{norouzian2019rain,norouzian2019experimental} suggest feasibility of operation at a range of up to $200$ m with the \textit{specific attenuation} (derivative of attenuation with respect to range) of \textasciitilde$10$ dB km$^{-1}$ over $0.1$-$0.3$ THz resulting in a path loss of \textasciitilde$4$ dB \cite{appleby2007millimeter}.
	
	For THz communications, high propagation losses and power limitations are compensated by the beamforming gains obtained through deployment of extremely dense nano-antenna arrays \cite{ummimoTareqOverview}, which may be based on graphene plasmonics \cite{ju2011graphene,xu2014design} or metasurfaces \cite{moccia2018suboptimal}. Analogous to the developments in mmWave communications, \cite{ummimoTareq} proposed a THz ultramassive multiple-input multiple-output (MIMO) architecture that employs an array-of-subarrays (AoSA) of nano-transceivers to increase the coverage and improve the data rates. Various follow-up works (see, e.g., \cite{ummimoTareqOverview} for an overview) showed further ultra-massive MIMO enhancements through waveform design, beamforming, and resource allocation.
	
	With this recent rise of both radar and communications applications at THz, it has been suggested \cite{chaccour2021seven} to integrate radar sensing and  communications functionalities in future wireless THz systems to facilitate spectrum sharing, enhance pencil beamforming, save hardware cost, and improve resource usage. This follows recent efforts in realizing such \textit{joint radar-communications} (JRC) architectures at mmWave \cite{mishra2019toward}, including for \textit{ultrashort ranges} \cite{duggal2020doppler}, joint MIMO-radar-MIMO-communications \cite{dokhanchi2020multi}, and distributed MIMO JRC \cite{liu2020codesigning}. In this paper, we focus on a ultra-massive MIMO structure for JRC at THz band.

	\begin{table*}[t]
		{\color{black}
			\caption{Comparison of mm-Wave and THz Transmission Characteristics}
			\label{tableComparison}
			\centering
			\begin{tabular}{p{0.11\textwidth}p{0.38\textwidth}p{0.44\textwidth}}
				\hline
				\hline
				\bf{Phenomenon}& \bf mmWave	& \bf THz \\
				\hline 
				Path Loss &The path loss exponent in (\ref{alphaLoS}) $\bar{\gamma}\approx 2$~\cite{teraMIMO}. Massive array structures are used to mitigate path loss~\cite{heath2016overview,mishra2019toward}. &The path loss exponent doubles, i.e., in (\ref{alphaLoS}) $\bar{\gamma}\approx 4$~\cite{teraMIMO}. Ultra-massive arrays are employed to mitigate path loss~\cite{teraMIMO,ummimoComMagYeLi,ummimoTareq}.  \\
				\hline
				Channel Model&Superposition of LoS and NLoS paths~\cite{heath2016overview,mimoHybridLeus1}. Channel estimation may solely exploit sparse reconstruction techniques e.g., compressed sensing (CS) and orthogonal matching pursuit (OMP)~\cite{heath2016overview}.& A dominant LoS path with
				multiple NLoS paths~\cite{teraMIMO,ummimoTareq,ummimoHBThzSVModel} as in (\ref{channelMatrixCSI}). Channel estimation may require joint OMP and true-time-delay (TTD) techniques~\cite{channelEstThz2,trueTimeDelayBeamSquint}. \\
				\hline
				Beam Alignment & Beams are \textit{squinted} but still cover the user across the entire bandwidth ~\cite{beamSquintGaoMagazine,beamSquintWang2019Nov}. This effect is largely dependent on the large number of antennas. Corrected via TTD processing at each antenna~\cite{trueTimeDelayBeamSquint,trueTimeDelayMagazine}.&Beams become totally \textit{split} and cannot cover the user with their mainlobes~\cite{thz_beamSplitConf,thz_beamSplit}. This split is a function of both bandwidth and array size. Corrected by delay-phase precoding (DPP)~\cite{thz_beamSplitConf,beamSplitDai2021Feb} or beam split phase correction as in (\ref{beamSplitEquation}). \\
				\hline
				Array Structure & Large arrays with fully-connected or subarray structures, where the latter has lower hardware complexity~\cite{hybridBFAltMin,heath2016overview}. & Much larger array structures with additional subarray levels to reduce the hardware complexity, e.g., AoSA~\cite{ummimoComMagYeLi}, widely-spaced multi-subarray (WSMS)~\cite{ummimoHBChallencesOpenProblems} or GoSA structures (Fig.~\ref{figHB}c).\\ 
				\hline
				\hline
			\end{tabular}
		}
	\end{table*}

	Analogous to their massive MIMO counterparts at mmWave, the ultra-massive MIMO architecture implies that fully digital beamforming is infeasible because of huge associated cost, area, and power. This necessitates the use of hybrid beamforming \cite{mimoHybridLeus1}, wherein the signal is  processed by both analog and digital beamformer. Although some recent works \cite{busari2019generalized,yuan2020hybrid,ning2021terahertz} investigate THz hybrid beamformer designs, they do not examine it in the context of the recently proposed, practically feasible ultra-massive MIMO, and more so, its application in THz JRC.
	
	{\color{black}While considering beamforming at THz, following unique characteristics (see also Table~\ref{tableComparison}) differentiate the THz-band from mmWave:
		\begin{enumerate}
			\item The path loss in THz channels includes both spreading loss and molecular absorption. The latter is more significant at THz than mmWave \cite{yuan2020hybrid}. \textcolor{black}{The severe path loss is compensated by deploying much larger antenna arrays that require more creative choices for subarray geometries.}
			\item Both line-of-sight (LoS) and non-LoS (NLoS) paths are significant at mmWave. However, at THz, the NLoS paths have insignificant contribution to the received power. This leads to LoS-dominant and NLoS-assisted communications scenario at \textcolor{black}{THz~\cite{faisal2020ultramassive,ummimoHBThzSVModel,Ning2021Jan,thz_beamSplit,trueTimeDelayBeamSquint}.}
			\item Significant attenuation implies shorter ranges for THz systems than mmWave~\cite{ummimoGrapheneArray}. \textcolor{black}{Additionally, the THz-specific molecular absorption effect leads to the channel bandwidth varying with the range \cite{hossain2019hierarchical}. This requires stricter limitations on the deployment and coverage of THz communications and radar.}
			\item \textcolor{black}{THz channels are \textit{extremely sparse} \cite{han2021hybrid} in the angular domain and have smaller angular spread than mmWave. Therefore, it is feasible to adopt subarrayed models such as AoSA~\cite{ummimoComMagYeLi} and GoSA to bring down the high-frequency hardware and computational complexities.} These subarrayed structures have been shown to overcome the limited communications range while also retaining a reasonable spatial multiplexing gain.
			\item THz channel exhibits peculiarities such as misalignment and phase uncertainties in phase shifters~\cite{teraMIMO}. The frequency-independent analog beamformers largely used in the broadband mmWave communications may lead to \textit{beam split} effect in THz channels: the generated beams split into different physical directions at each subcarrier due to ultra-wide bandwidth and large number of antennas~\cite{thz_beamSplit}. \textcolor{black}{This phenomenon has also been called as \emph{beam squint} in mm-Wave works~\cite{beamSquintGaoMagazine,beamSquintWang2019Nov}. While both beam squint and beam split pertain to a similar phenomenon, the latter has more severe achievable rate degradation. In particular, the main lobes of the array gain corresponding to the lowest and highest subcarrier frequencies do not overlap at THz at all while there is a relatively small deviation in mmWave band (see Fig.~\ref{fig_BeamSplit}) \cite{beamSplitDai2021Feb,thz_beamSplitConf}. While the beam-squint depends on the array size, the beam split is a function of both wide bandwidth and large arrays \cite{beamSplitDai2021Feb}.}
			\item The THz-band has several other propagation and scattering effects \cite{teraMIMO}. The specular scattering is less dominant and partially-diffused scattering is a more appropriate model for THz. The THz channel coherence time may be smaller than the symbol time and the channel may no longer be considered time-invariant. Further, the molecular absorption is range-dependent leading to very different noise models than mmWave. At much shorter distances, the THz bandwidth is also range-dependent and spherical wave propagation must be accounted for.
		\end{enumerate}

	}

\begin{figure*}[t]
		\centering
		{\includegraphics[draft=false,width=\textwidth]{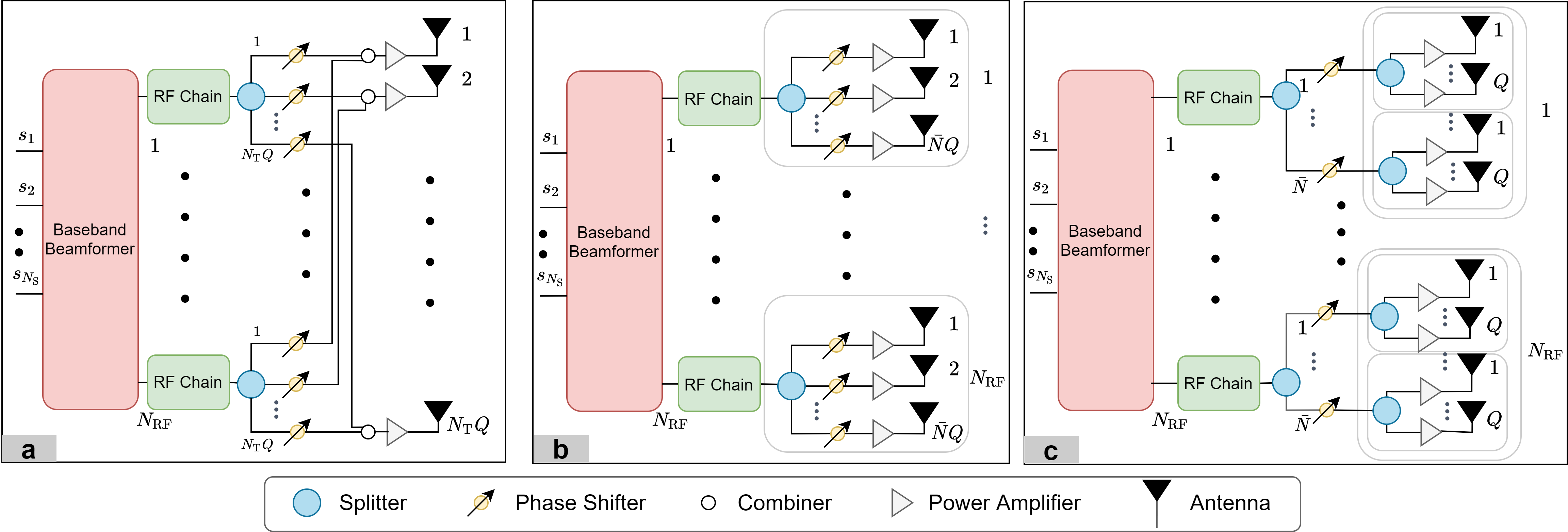} }
		\caption{Hybrid beamforming-based transmitter structures for (a) fully connected, (b) partially-connected array of subarrays (AoSAs) and (c) partially-connected groups of subarrays (GoSAs) architectures. While all the architectures employ $N_\mathrm{T}Q$ antennas with $N_\mathrm{RF}$ RF chains, each antenna is connected to each RF chain via combiners in the fully connected model with $N_\mathrm{T}QN_\mathrm{RF}$ phase-shifters. In partially-connected AoSA, the same RF chain is connected to $\bar{N}Q$ ($\bar{N} = \frac{N_\mathrm{T}}{N_\mathrm{RF}}$) antennas with $N_\mathrm{T}Q$ phase-shifters totally. In partially-connected GoSA model, each RF chain is connected to $\bar{N}Q$ antennas while each phase-shifter is connected only $Q$ antennas, introducing $N_\mathrm{T}$ group of subarrays with only  $N_\mathrm{T}$ phase-shifters.}
		\label{figHB}
	\end{figure*}

	\textcolor{black}{In this work, we consider the THz hybrid beamforming problem by incorporating the above-mentioned unique THz features except the last scattering characteristic, which is beyond the scope of this paper.} Contrary to prior works, we focus on the THz \textcolor{black}{wideband} hybrid beamforming for a ultra-massive MIMO JRC configuration. To this end, we develop both model-based and model-free techniques that rely on both channel state information (CSI) and channel covariance matrix. 
	To reduce the hardware complexity, we propose a group of subarrays (GoSAs) structure, in which the antenna elements in the same subarray are connected to the same phase-shifter. \textcolor{black}{In other words, GoSA forms an array of subarray-of-subarrays, which is different than the prior array-of-subarray (AoSA) structures (see, e.g., Fig.~\ref{figHB}).}  Thus, the proposed structure employs even fewer phase-shifters than that of fully-connected arrays or partially-connected  AoSA structures~\cite{ummimoComMagYeLi,ummimoGrapheneArray}, while providing satisfactory radar and communications performance in terms of the beampattern and the spectral efficiency, respectively. In order to improve the radar performance, the higher degrees of freedom (DoF) are provided by using partially-connected GoSAs. Nevertheless, partially-connected structure has poor spectral efficiency performance compared to the fully-connected array. Hence, we suggest a partially-connected with overlapped (PCO) GoSA structure for performance improvement. To design the hybrid beamformers based on the PCO structure, we propose a modified version of the manifold optimization (MO)-based alternating minimization (AltMin) technique~\cite{hybridBFAltMin}, which is originally suggested to solve the beamformer design problem in fully-connected arrays. Our numerical experiments show that the proposed approach has much lower hardware complexity than the state-of-the-art techniques, while maintaining satisfactory radar and communications performance. In this work, our main contributions are:\\
	\textbf{1) THz ultra-massive MIMO JRC.} Our proposed JRC approach based on ultra-massive MIMO is inspired by recent advancements in THz technologies and is, therefore, closer to practical feasibility. It is particularly helpful for short-range vehicular applications, wherein the ego vehicle simultaneously communicates with the user equipment and detect/track the radar targets with higher angular resolution due to high beamforming gain of using ultra-massive number of antennas. \\
	\textbf{2) Model-based THz hybrid beamforming.} Previous research \cite{busari2019generalized,yuan2020hybrid,ning2021terahertz} examined THz hybrid beamforming without ultra-massive MIMO. Our optimization-based hybrid beamforming for ultra-massive MIMO relies on both CSI and channel covariance matrix. While the former provides higher spectral efficiency, the latter has lower channel overhead at the cost of slight performance loss.\\
	\textbf{3) Novel GoSA structure.} We propose GoSA structure to lower the hardware cost which could be high for THz systems due to use of a large number of antennas. GoSA allows us to employ fewer number of phase shifters as compared to AoSA. We analyze the performance of GoSA with respect to several design parameters, such as the number of antennas and the antenna spacing. To provide a trade-off between the hardware complexity and the spectral efficiency, PCO-based analog precoder is proposed based on modified manifold optimization method.\\
	{\color{black}\textbf{4) Beam split correction.} We present a hardware-efficient approach to correct the beam split effect in THz channels arising from their ultra-wide bandwidth. While prior works~\cite{thz_beamSplit,thz_beamSplitConf} consider an additional time-delay network for this operation, the proposed approach effectively mitigates the beam split effect without requiring such a complex structure.} \\
	\textbf{5) Deep learning (DL) solutions.} We design two learning models using convolutional neural networks (CNNs), one of which is employed to estimate the direction of the radar targets, whereas the other is used to design the hybrid beamformers. \textcolor{black}{While DL-based beamforming techniques~\cite{elbirQuantizedCNN2019,elbir2019online} have been proposed earlier, THz JRC hybrid beamformer design remains unexamined in prior literature.}  
	
	Throughout this paper, we denote the vectors and matrices by boldface lower and upper case symbols, respectively. In case of a vector $\mathbf{a}$, $[\mathbf{a}]_{i}$ represents its $i$-th element. For a matrix $\mathbf{A}$, $[\mathbf{A}]_{i,j}$ denotes the $(i,j)$-th entry. The $\mathbf{I}_N$ is the identity matrix of size $N\times N$; $\mathbb{E}\{\cdot\}$ denotes the statistical expectation; $\textrm{rank}(\cdot)$ denotes the rank of its matrix argument; $\|\cdot\|_\mathcal{F}$ is the Frobenius norm; $(\cdot)^{\dagger}$ denotes the Moore-Penrose pseudo-inverse; and $\angle\{\cdot\}$ denotes the angle of a complex scalar/vector. The Kronecker and element-wise Hadamard product are denoted by $\otimes$ and  $\odot$,  respectively. The notation expressing a convolutional layer with $N$ filters/channels of size $D\times D$ is given by  $N$@$ D\times D$.

	The rest of the paper is organized as follows. In the next section, we describe the system and channel models of GoSA-based ultra-massive MIMO JRC and formulate the beamformer design problem. Section~\ref{sec:modelBased} introduces the CSI- and channel covariance matrix-based beamformer solutions along with extension to broadband beamforming. We follow this in Section~\ref{sec:modelFree} by DL-based solution. We validate our models and methods through numerical experiments in Section~\ref{sec:numexp} and conclude in Section~\ref{sec:summ}.
	
	\section{System Model and Problem Formulation}
	\label{sec:sysmod}
	{\color{black}We consider a wideband ultra-massive MIMO architecture} in the context of a JRC system for a vehicle to vehicle (V2V) and vehicle to device (V2D) scenario, in which the transmitter (TX) senses the environment via probing waveforms to the targets and communicates with the receiver (RX), as illustrated in Fig.~\ref{figHB_TX}. The antenna arrays at the TX and the RX employ graphene-based plasmonic nano-antennas, which are placed on a metallic surface layer, with a dielectric layer between them~\cite{ummimoGrapheneArray,ummimoTareq,ummimoComMagYeLi}. The antennas form GoSA structure as each subarray consists of $Q_\mathrm{x}\times Q_\mathrm{y}$ uniform rectangular arrays (URAs) with $Q=Q_\mathrm{x}Q_\mathrm{y}$ antennas, as shown in Fig.~\ref{figHB_TX}. Also, there are $N_\mathrm{T} = N_\mathrm{T_x} N_\mathrm{T_y}$ and $N_\mathrm{R}=N_\mathrm{R_x}N_\mathrm{R_y}$ subarrays of size $Q$ at the TX and RX, respectively, which form an $N_\mathrm{T}Q\times N_\mathrm{R}Q$ ultra-massive MIMO transceiver architecture. In each $Q_\mathrm{x}\times Q_\mathrm{y}$ subarray, the antenna spacing along the $x$- and $y$-axis are $ \delta_\mathrm{x}, \delta_\mathrm{y}$ and the distance between each subarray are $\Delta_\mathrm{x},\Delta_\mathrm{y}$, respectively.

	\subsection{Communications Model}
	
	{\color{black}In the downlink, the TX with $N_\mathrm{T}$ subarrays, each of which has $Q$ antenna elements,  aims to transmit $N_\mathrm{S}$ data streams towards the RX in the form of $\mathbf{s}[m] = [s_1[m],\dots,s_{N_\mathrm{S}}[m]]^\textsf{T}$ by using hybrid analog and digital beamformers with $N_\mathrm{RF}$ RF chains, where $\mathbb{E}\{\mathbf{s}[m]\mathbf{s}^\textsf{H}[m]\} = \mathbf{I}_{N_\mathrm{S}}$ and $N_\mathrm{S}\leq N_\mathrm{RF}$. Here, $m \in  \mathcal{M} = \{1,\dots, M\}$ and $M$ is the number of subcarriers.} Due to beamforming at subarray level, each subarray of size $Q$ generates a single beam~\cite{ummimoTareq}. This is done by connecting the $Q$ antennas in each subarray to a single phase-shifter to lower the hardware complexity. {\color{black}Thus, the TX first applies subcarrier-dependent $N_\mathrm{RF}\times N_\mathrm{S}$ baseband precoder $\mathbf{F}_\mathrm{BB}[m]$. The signal is, then, transformed to the time-domain via $M$-point inverse fast Fourier transform	(IFFT). After adding the cyclic prefix, the TX employs
		a subcarrier-independent RF precoder $\mathbf{F}_\mathrm{RF}\in \mathbb{C}^{N_\mathrm{T}\times N_\mathrm{RF}}$ by employing $N_\mathrm{T}$ phase-shifters, as shown in Fig.~\ref{figHB}c.} In conventional fully-connected structures (see Fig.~\ref{figHB}a), each antenna is connected to $N_\mathrm{RF}$ RF chains while the AoSA model in Fig.~\ref{figHB}b has a partially-connected structure and it connects each RF chain to $\bar{N}Q$ antennas in each subarray, where $\bar{N} = \frac{N_\mathrm{T}}{N_\mathrm{RF}}$~\cite{ummimoComMagYeLi}. In this work, we propose a GoSA architecture, as shown in Fig.~\ref{figHB}c, in which $N_\mathrm{T}Q$ antennas are partitioned into $N_\mathrm{RF}$ groups, each of which has $\bar{N}Q$ antennas. Also, each group consists of $\bar{N}$ subarrays of size $Q$. \textcolor{black}{The main difference between AoSA and GoSA is that each RF chain is connected to $\bar{N}Q$ phase-shifters in the former while the each RF chain is connected to only $\bar{N}$ phase-shifters in the latter. Hence, the number of phase-shifters in GoSA is $Q$ times lower than that of AoSA.} In GoSA, we assume that the antennas in each subarray are fed with the same phase shift to reduce the hardware complexity and power consumption, which is critical in THz systems.
	
	In the proposed GoSA model, the RF precoder has unit-modulus constraints, i.e., $|[\mathbf{F}_\mathrm{RF}]_{i,j}| = \frac{1}{\sqrt{N_\mathrm{T}}}$ as $i \in \{1,\dots, N_\mathrm{T}\}$ and $j \in \{1,\dots, N_\mathrm{RF}\}$, since $\mathbf{F}_\mathrm{RF}$ is constructed by using phase-shifters. Furthermore, we have power constrained $\sum_{m \in \mathcal{M}}\|\mathbf{F}_\mathrm{RF}\mathbf{F}_\mathrm{BB} [m] \|_{\mathcal{F}} = MN_\mathrm{S}$. Thus, the $N_\mathrm{T}\times 1$ transmitted signal from the TX is given by $		\mathbf{x}[m] = \mathbf{F}_\mathrm{RF}\mathbf{F}_\mathrm{BB}[m]\mathbf{s}[m].$
	%


	{\color{black}Assuming frequency-selective fading over multi-carrier transmission between the TX and  RX~\cite{ummimoTareq}, the received signal at the RX is given by
		\begin{align}
		\label{receivedSignal}
		\mathbf{y}[m] = \sqrt{\rho} \mathbf{H}[m] \mathbf{F}_\mathrm{RF} \mathbf{F}_\mathrm{BB}[m]\mathbf{s}[m] + \mathbf{n}[m],
		\end{align}
		where $\mathbf{y}[m]\in \mathbb{C}^{N_\mathrm{R}}$ is the output of $N_\mathrm{R}$ subarrayed antennas at the RX, $\rho$ is the received power and $\mathbf{n}[m]\in \mathbb{C}^{N_\mathrm{R}}$ denotes the additive white Gaussian noise (AWGN) vector with $\mathbf{n}[m]\sim \mathcal{CN}(\mathbf{0}, \sigma_n^2\mathbf{I}_{N_\mathrm{R}})$. $\mathbf{H}[m]\in \mathbb{C}^{N_\mathrm{R}\times N_\mathrm{T}}$ denotes the THz channel matrix between the TX and the RX. }

	{\color{black}In THz transmission, the wireless channel $\mathbf{H}[m]$  can be represented by a single dominant  LoS path with assisting a few NLoS paths, which are small due to large reflection loses, scattering and refraction~\cite{ummimoComMagYeLi,ummimoTareq,teraMIMO}. Channel modeling at THz-band has been a challenge largely because of lack of realistic measurements. \textcolor{black}{Very recently, measurement campaigns at 140 GHz have been reported \cite{xing2018propagation,ju2021140}. In particular, \cite{ju2021140} states that, while the delay/angular spread at 140 GHz and lower frequencies are comparable, the correlation distance of shadow fading at the former is much shorter. The same study mentions multiple NLoS and dominant LoS paths at low-THz. This work is, however, not focused on only 140 GHz, and, therefore, employs the assumptions theorized for the entire upper-mmWave/low-THz region as in \cite{teraMIMO}.}
		
		While the ray-tracing techniques assume the channel to be sparse and dominated by the LoS component for the graphene nano-transceivers \cite{ummimoTareqOverview}, the other channel models such as the 3GPP model \cite{3gppChannelModelThz,umMIMO3gpp_IRS} are also popular for THz beamforming. In this work, we adopt the Saleh-Valenzuela (SV) THz channel model channel~\cite{ummimoHBThzSVModel,ummimoTareq}, wherein $\mathbf{H}[m]$ is constructed by the superposition of a single LoS path and the contribution of $N_\mathrm{clu}$ cluster of $N_\mathrm{ray}$ NLoS paths. Then, the $N_\mathrm{R}\times N_\mathrm{T}$ THz ultra-massive MIMO channel matrix is given by~\cite{teraMIMO}
		\begin{align}
		\label{channelMatrixCSI}
		&\mathbf{H}[m] = \nonumber \\
		&  \hspace{-2pt}\gamma \big( \alpha_{1,m}  \mathbf{A}_\mathrm{R}^m(\Theta_{1}) \mathbf{A}_\mathrm{T}^{m^\textsf{H}}(\Psi_1)  \hspace{-2pt}  + \hspace{-2pt}
		\sum_{l=2}^{L} \hspace{-2pt} \alpha_{l,m} \mathbf{A}_\mathrm{R}^m(\Theta_{l})\mathbf{A}_\mathrm{T}^{m^\textsf{H}}(\Psi_{l}) \big),
		\end{align}
		where $\gamma = \sqrt{\frac{N_\mathrm{T}N_\mathrm{R}}{L}}$ and $L = 1 + N_\mathrm{clu}N_\mathrm{ray}$ denotes the total number of LoS and NLoS paths. Furthermore, $\alpha_{l,m}$ represents the channel gain of the $l$th path for $m$th subcarrier, and we have 
		\begin{align}
		\alpha_{l,m} = \left\{\begin{array}{ll}  \alpha_m^\mathrm{LoS}, & l=1 \\
		\alpha_{rc,m}^\mathrm{NLoS}, & l>1, \hspace{2pt} l= (N_\mathrm{ray}-1)c + r  +1
		\end{array}\right.,
		\end{align}
		for which $l \in \{1,\dots, L\}$, $r\in \{1,\dots, N_\mathrm{ray}\}$ and $c \in \{1,\dots, N_\mathrm{clu}\}$.  $\alpha_m^\mathrm{LoS}\in \mathbb{C}$ denotes channel gain of the LoS path and it is defined as
		\begin{align}
		\label{alphaLoS}
		\alpha_m^{\mathrm{LoS}} = (\frac{c_0}{4 \pi f_m \bar{d} })^{\frac{\bar{\gamma}}{2}} e^{-\frac{1}{2} \bar{\kappa}(f_m) \bar{d}},
		\end{align} where $c_0$ is the speed-of-light, $\bar{\kappa}(f_m)$ is frequency-dependent molecular absorption coefficient and $\bar{d}$ is the distance between the TX and RX. In mmWave channels, the path loss exponent $\bar{\gamma}$ is around $2$ while it has typical values between $3$ and $4.5$ in THz channels for dense urban environments~\cite{teraMIMO}. 	 $\alpha^\mathrm{NLoS}_{m,rc}= |\alpha^\mathrm{NLoS}_{m,rc}| e^{j\bar{\beta}_{m,rc}} $ corresponds to the channel gain of the $r$th NLoS path in the $c$th cluster and $\bar{\beta}_{m,rc}$ is an independent uniformly distributed phase shift  over $[0,2\pi)$~\cite{teraMIMO}. The notations $\Theta_{l} = \{\phi_{l},\theta_{l}\}$ and $\Psi_{l} = \{\varphi_{l},\vartheta_{l}\}$ denote the azimuth/elevation angle-of-arrival (AoA) and angle-of-departure (AoD) of the received/transmitted paths at the RX and the TX, respectively. The matrices $\mathbf{A}_\mathrm{R}^m(\Theta_{l})\in\mathbb{C}^{N_\mathrm{R}\times Q}$ and $\mathbf{A}_\mathrm{T}^m(\Psi_{l})\in \mathbb{C}^{N_\mathrm{T}\times Q}$ are the steering matrices corresponding to the AoAs and AoDs of the GoSAs, respectively, and they are defined as
		\begin{align}
		\mathbf{A}_\mathrm{R}^m(\Theta_{l}) = \left[ \hspace{-3pt}\begin{array}{c}
		\mathbf{a}_{\mathrm{R},1}^{m^\textsf{T}}(\Theta_{l}) \\
		\vdots \\ 
		\mathbf{a}_{\mathrm{R},N_\mathrm{R}}^{m^\textsf{T}}(\Theta_{l}) \end{array} \hspace{-3pt}\right],
		\mathbf{A}_\mathrm{T}^m(\Psi_{l}) = \left[ \hspace{-3pt}\begin{array}{c}\mathbf{a}_{\mathrm{T},1}^{m^\textsf{T}}(\Psi_{l}) \\
		\vdots\\
		\mathbf{a}_{\mathrm{T},N_\mathrm{T}}^{m^\textsf{T}}(\Psi_{l}) \end{array} \hspace{-3pt}\right]
		\end{align}
		where $\mathbf{a}_{\mathrm{R},n_R}^m(\Theta_{l})$ ($\mathbf{a}_{\mathrm{T},n_T}^m(\Psi_{l}) $) is $Q\times 1$ steering vector corresponding to the antennas in the $n_R$th ($n_T$th) subarray for $n_R\in \{1\dots, N_\mathrm{R}\}$ ($n_T\in \{1,\dots, N_\mathrm{T}\}$), respectively.
		The $i$th element of the transmit steering vector $\mathbf{a}_{\mathrm{T},n_T}^m(\Psi_{l})$ is given by 
		\begin{align}
		\label{svector}
		[\mathbf{a}_{\mathrm{T},n_T}^m(\Psi_{l})]_i =   \frac{1}{\sqrt{N_\mathrm{T}}}\exp \{-j\frac{2\pi }{\lambda_m} \boldsymbol{\kappa}_{n_T,i}^\textsf{T} \boldsymbol{\Omega}_{l}   \},
		\end{align}
		where $\lambda_m = \frac{c_0}{f_m}$ is the wavelength for the subcarrier $m$ with frequency of $f_m = f_c + \frac{B}{M}(m - 1 - \frac{M-1}{2})$, where $B$ denotes the bandwidth.  $\boldsymbol{\kappa}_{n_T,i} = [x_{n_T,i},y_{n_T,i},z_{n_T,i}]^\textsf{T}$ denotes the position of the $i$th antenna of the $n_T$th subarray in Cartesian coordinate system and $\boldsymbol{\Omega}_{l}$ is a direction-dependent parameter defined as $	\boldsymbol{\Omega}_{l} =  [\cos\varphi_{l}\sin\vartheta_{l}, \sin\varphi_{l}\sin\vartheta_{l}, \cos\vartheta_{l}]^\textsf{T}.$
		The structure of  $\mathbf{a}_\mathrm{R}^m(\Theta_{l})$  is similar to that of $\mathbf{a}_\mathrm{T}^m(\Psi_{l})$.  Without loss of generality, we assume that the antennas are perfectly calibrated against mutual coupling and gain/phase mismatches~\cite{ummimoTareq}. 
		Finally, by exploiting the GoSA structure, the $(n_R,n_T)$th element of $\mathbf{H}[m]\in N_\mathrm{R}\times N_\mathrm{T}$ is given as
		\begin{align}
		[\mathbf{H}[m]]_{n_R,n_T} = \gamma \sum_{l = 1}^{L} \alpha_{l,m}   \mathbf{a}_{\mathrm{R},n_R}^m(\Theta_{l})\odot \mathbf{a}_{\mathrm{T},n_T}^{m^*}(\Psi_{l}).
		\end{align}
		
	}
	
	By connecting the $Q$ antennas in the subarrays to a single phase-shifter, we are able to construct an $N_\mathrm{T}\times N_\mathrm{RF}$, (instead of $N_\mathrm{T}Q\times N_\mathrm{RF}$ as in Fig.~\ref{figHB}a and Fig.~\ref{figHB}b) RF precoder, as illustrated in Fig.~\ref{figHB}c. Using partially-connected GoSA, the associated RF precoder has the form of
	\begin{align}
	\label{Frf}
	\mathbf{F}_\mathrm{RF} = \left[\begin{array}{cccc}
	{\mathbf{u}}_1 & \mathbf{0} & \cdots & \mathbf{0} \\
	\mathbf{0} & {\mathbf{u}}_2 & \cdots & \mathbf{0} \\
	\vdots & \mathbf{0} & \ddots & \mathbf{0} \\
	\mathbf{0} & \mathbf{0} & \cdots & {\mathbf{u}}_{N_\mathrm{RF}}
	\end{array} \right]\in \mathbb{C}^{N_\mathrm{T}\times N_\mathrm{RF}},
	\end{align}
	where ${\mathbf{u}}_i\in \mathbb{C}^{\bar{N}}$ represents a portion $N_\mathrm{T}\times 1$ phase-shifter values with indices $\{(i-1)\bar{N}+1,\dots, i\bar{N}\}$ for $i \in \{ 1,\dots, N_\mathrm{RF}\}$, where $\bar{N} = \frac{N_\mathrm{T}}{N_\mathrm{RF}}$. Each entry of $\mathbf{u}_i$ is then applied to $Q$ antennas in $N_\mathrm{T}$ subarrays to steer the transmitted beams (see, e.g., Fig.~\ref{figHB}c) so that a total of $N_\mathrm{T}Q$ antennas are fed.

	To address the performance degradation due to GoSA, the columns of $\mathbf{F}_\mathrm{RF}$ is designed with overlapping terms~\cite{phasedArrayMIMOradar,overlappedSubarrayMassiveMIMO}. Assume $\bar{\mathbf{u}}_i \in \mathbb{C}^{\bar{M}}$ to include the overlapped phase-shifter terms, where $\bar{M} \in [\bar{N}, N_\mathrm{T} - N_\mathrm{RF}+1]$, for which $\bar{M}= \bar{N}$ provides non-overlapped partially-connected structure as in (\ref{Frf}) while $\bar{M} = N_\mathrm{T} - N_\mathrm{RF}+1$ provides maximum overlap among the phase-shifters. In this case, the performance improvement is at the cost of using more phase-shifters. Nevertheless, it still has lower number of phase-shifters as compared to the partially non-overlapped case in conventional AoSA (see, e.g., Fig.~\ref{fig_NumPhaseShift}). The use of partially-connected/PCO GoSA structure provides higher DoF as compared to the simple phased-array MIMO radar structure, for which $N_\mathrm{RF} = 1$~\cite{phasedArrayMIMOradar} and we have a fully-connected MIMO structure when $N_\mathrm{RF} = N_\mathrm{T}$.  While MIMO radar outperforms the phased-array in terms of angular resolution and DoF for parameter estimation and parameter identification, phase-array provides higher coherent processing gain and lower computation and hardware complexity~\cite{radarCommSurvey}. This complexity is further reduced by using the GoSA structure by feeding each of $Q$ antennas with the same phase shift.	Thus, the partially-connected GoSA provides a trade-off between the DoF and the hardware complexity, both of which increase as $N_\mathrm{RF} \rightarrow N_\mathrm{T}$.
	
	In communications-only systems, the aim is to design the hybrid precoders such that the spectral efficiency at the TX is maximized~\cite{elbirQuantizedCNN2019,heath2016overview,hybridBFAltMin}, while there are also other related performance metrics, such as energy-efficiency~\cite{energyEfficencyHB} and minimum mean-squared-error (MMSE)~\cite{mmseHB}. By decoupling the beamformer design problem at the TX and the RX, the mutual information at the TX is maximized instead of spectral efficiency, for which  a perfect combiner is assumed at the receiver~\cite{heath2016overview}. Once the transmitter is designed, the receive beamforming design is done by using the MMSE as performance metric as in~\cite{heath2016overview,hybridBFAltMin,elbirQuantizedCNN2019}. Then the mutual information of the communications system is given by {\color{black}
		\begin{align}
		\label{def:R}
		\mathcal{I} = \frac{1}{M}\sum_{m = 1}^{M} \mathcal{I}(\mathbf{F}_\mathrm{RF},\mathbf{F}_\mathrm{BB}[m] ),
		\end{align}
		where 
		\begin{align}
		\mathcal{I}(\mathbf{F}_\mathrm{RF},\mathbf{F}_\mathrm{BB}[m] ) = &\log_2  \bigg|\mathbf{I}_{N_\mathrm{R}} + \frac{\rho}{N_\mathrm{S}\sigma_n^2} \mathbf{H} [m] \mathbf{F}_\mathrm{RF}\mathbf{F}_\mathrm{BB}[m]   \nonumber\\
		&\times \mathbf{F}_\mathrm{BB}^\textsf{H}[m]\mathbf{F}_\mathrm{RF}^\textsf{H}\mathbf{H}^\textsf{H}[m] \bigg|,
		\end{align}
		corresponds to the mutual information for subcarrier $m$. 	We note here that the maximization of (\ref{def:R}) is provided by exploiting the similarity between the hybrid beamformer $\mathbf{F}_\mathrm{RF}\mathbf{F}_\mathrm{BB} [m]$ and the optimal unconstrained beamformer $\mathbf{F}_{\mathrm{C}} [m]\in \mathbb{C}^{N_\mathrm{T}\times N_\mathrm{S}}$. The latter is obtained from the right singular matrix of the channel matrix $\mathbf{H} [m]$ \cite{hybridBFAltMin,heath2016overview}. The singular value decomposition of the channel matrix is  $\mathbf{H} [m] = \mathbf{U}_{\mathbf{H}} [m] \boldsymbol{\Pi} [m] \mathbf{V}_{\mathbf{H}}^\textsf{H} [m]$, where $\mathbf{U}_{\mathbf{H}} [m]\in \mathbb{C}^{N_\mathrm{R}\times \mathrm{rank}(\mathbf{H} [m])}$ and $\mathbf{V}_{\mathbf{H}} [m]\in \mathbb{C}^{N_\mathrm{T} \times \mathrm{rank}(\mathbf{H} [m])}$ are the left and the right singular value matrices of the channel matrix, respectively,  and $\boldsymbol{\Pi} [m]$ is $\mathrm{rank}(\mathbf{H} [m])\times \mathrm{rank}(\mathbf{H} [m])$ matrix composed of the singular values of $\mathbf{H} [m]$ in descending order. By decomposing $\boldsymbol{\Pi} [m]$ and $\mathbf{V}_{\mathbf{H}} [m]$ as $\boldsymbol{\Pi} [m] = \mathrm{diag}\{ \widetilde{\boldsymbol{\Pi}} [m],\overline{\boldsymbol{\Pi}} [m] \},\hspace{5pt} \mathbf{V}_{\mathbf{H}} [m] = [\widetilde{\mathbf{V}}_{\mathbf{H}} [m],\overline{\mathbf{V}}_{\mathbf{H}} [m]],$
		where $\widetilde{\mathbf{V}}_{\mathbf{H}} [m]\in \mathbb{C}^{N_\mathrm{T}\times N_\mathrm{S}}$, the unconstrained precoder is readily obtained as $\mathbf{F}_{\mathrm{C}} [m] = \widetilde{\mathbf{V}}_{\mathbf{H}} [m]$ \cite{heath2016overview}. Then, the maximization of (\ref{def:R}) is achieved by minimizing the Euclidean distance between $\mathbf{F}_\mathrm{C} [m]$ and $\mathbf{F}_\mathrm{RF}\mathbf{F}_\mathrm{BB} [m] $ as
		\begin{align}
		\label{problemCom1}
		\minimize_{\mathbf{F}_\mathrm{RF},\{\mathbf{F}_\mathrm{BB}[m]\}_{m \in \mathcal{M}}}  &\hspace{3pt} \frac{1}{M}\sum_{m \in \mathcal{M}}\|\mathbf{F}_\mathrm{RF}\mathbf{F}_\mathrm{BB}[m]  -  \mathbf{F}_\mathrm{C}[m]\|_\mathcal{F} \nonumber \\
		\subjectto & \sum_{m \in \mathcal{M}}\| \mathbf{F}_\mathrm{RF}\mathbf{F}_\mathrm{BB}[m] \|_\mathcal{F} = MN_\mathrm{S}, \nonumber \\
		& |[\mathbf{F}_\mathrm{RF}]_{i,j}| = \frac{1}{\sqrt{N_\mathrm{T}}}, \hspace{5pt} \forall i,j.
		\end{align}
	}

	\subsection{Radar Model}
	The goal of radar processing is to achieve the highest possible SNR gain towards the direction of interest. The radar first transmits an omni-directional waveforms  to detect the unknown targets within the angular space of interest in the search phase, then it generates directional beams towards to the targets for tracking purposes~\cite{radarCommSurvey}. We assume a subarrayed MIMO radar architecture with GoSAs, wherein each GoSA is used to coherently transmit waveforms that are orthogonal to the ones generated by other GoSAs~\cite{phasedArrayMIMOradar}, thereby, coherent processing gain is achieved. To this end, the transmit waveform of the $k$th GoSA ($k\in \{1,\dots, K\}$) is designed as $w_k(t) = W(t)e^{j2\pi k \Delta_ft}$, $0<t<T_0$, where $W(t)$ is the pulse shape with duration of $T_0$, so that the orthogonality of $w_k(t)$ is satisfied for a variety of time delays and Doppler shifts if the frequency increment among the GoSA waveforms satisfies $\Delta_f = | f_{k+1} - f_k | \gg 1/T_0$~\cite{phasedArrayMIMOradar}. Denote $\{\Phi_1,\dots, \Phi_K\}$ as the set of target directions ($\Phi_k = (\bar{\varphi}_k,\bar{\vartheta}_k)$), then, the $N_\mathrm{T}\times K$ GoSA-MIMO radar-only beamformer is modeled as $\mathbf{F}_\mathrm{R} = \mathrm{blkdiag}\{	{\mathbf{v}}_1,\dots, 	{\mathbf{v}}_K\}$ similar to (\ref{Frf}),
	where $\mathbf{v}_k\in \mathbb{C}^{\bar{K}}$ denotes the values of the transmit steering vector $\mathbf{a}_\mathrm{T}(\Phi_k)\in \mathbb{C}^{N_\mathrm{T}}$  with indices $\{(k-1)\bar{K}+1,\dots, k\bar{K}\}$ for $k = 1\dots, K$ and $\bar{K} = \frac{N_\mathrm{T}}{K}$. It is possible to construct $\mathbf{F}_\mathrm{R}$ via overlapped GoSA with $\bar{\mathbf{v}}_k\in \mathbb{C}^{N_\mathrm{T} - K +1}$ for $k \in \{1,\dots,K\}$.
	
	%
	The estimation of the target directions $\{\Phi_k\}_{k=1}^K$ is performed in the search phase of the radar. This is achieved via both: model-based methods, such as MUSIC (\textit{mu}ltiple \textit{si}gnal \textit{c}lassification) algorithm~\cite{music}, and model-free techniques based on DL~\cite{elbir_DL_MUSIC}. In this work, we assume that search operation is completed and the direction information of the targets is acquired prior to the beamformer design. The beampattern of the radar with GoSA structure is 
	\begin{align}
	B(\tilde{\Phi},m) = \mathrm{Trace}\{\mathbf{A}_\mathrm{T}^\textsf{H}(\tilde{\Phi})\mathbf{R}[m]\mathbf{A}_\mathrm{T}(\tilde{\Phi})\},
	\end{align}
	where 	\textcolor{black}{$\mathbf{R}[m]\in \mathbb{C}^{N_\mathrm{T}\times N_\mathrm{T}}$ }is the covariance matrix of the transmitted signal, then the design of the radar beampattern is equivalent to the design of the covariance matrix of the radar probing signals subject to the hybrid architecture of the beamformers.	In case of a single target scenario, the optimal beamformer is known to be conventional nonadaptive beamformer, i.e., steering vector corresponding to the direction of interest~\cite{phasedArrayMIMOradar}. When there are multiple targets, the covariance matrix of the transmitted signal is utilized.	In case of multiple targets in radar-only scenario with hybrid beamforming, we define the covariance matrix of the transmitted signal {\color{black}$\mathbf{x}[m]$ as
		\begin{align}
		\mathbf{R}[m] &= \mathbb{E}\{\mathbf{x}[m]\mathbf{x}^\textsf{H}[m]  \}\nonumber \\ &= \mathbb{E}\{\mathbf{F}_\mathrm{RF}\mathbf{F}_\mathrm{BB}[m]\mathbf{s}[m]\mathbf{s}^\textsf{H}[m]\mathbf{F}_\mathrm{BB}^\textsf{H}[m]\mathbf{F}_\mathrm{RF}^\textsf{H}  \}, \nonumber \\
		&= \mathbf{F}_\mathrm{RF}\mathbf{F}_\mathrm{BB}[m]\mathbb{E}\{\mathbf{s}[m]\mathbf{s}^\textsf{H}[m]\}\mathbf{F}_\mathrm{BB}^\textsf{H}[m]\mathbf{F}_\mathrm{RF}^\textsf{H},\nonumber\\
		&= \mathbf{F}_\mathrm{RF}\mathbf{F}_\mathrm{BB}[m]\mathbf{F}_\mathrm{BB}^\textsf{H}[m]\mathbf{F}_\mathrm{RF}^\textsf{H},
		\end{align}
		which requires the design of hybrid beamformers $\mathbf{F}_\mathrm{RF}$ and $\mathbf{F}_\mathrm{BB}[m]$. The hybrid beamformer design problem for radar-only system is solved by minimizing the Euclidean distance between $\mathbf{F}_\mathrm{RF}\mathbf{F}_\mathrm{BB}[m]$ and $\mathbf{F}_\mathrm{R}\mathbf{P}[m]$ as
		\begin{align}
		\label{problemRadar1}
		\minimize_{\mathbf{F}_\mathrm{RF},\{\mathbf{F}_\mathrm{BB}[m],\mathbf{P}[m]\}_{m\in \mathcal{M}}} &\hspace{1pt} \frac{1}{M}\sum_{m \in \mathcal{M}}\|\mathbf{F}_\mathrm{RF}\mathbf{F}_\mathrm{BB}[m]  -  \mathbf{F}_\mathrm{R}\mathbf{P}[m]\|_\mathcal{F} \nonumber \\
		\subjectto & \sum_{m \in \mathcal{M}}\| \mathbf{F}_\mathrm{RF}\mathbf{F}_\mathrm{BB}[m] \|_\mathcal{F} = MN_\mathrm{S}, \nonumber \\
		& |[\mathbf{F}_\mathrm{RF}]_{i,j}| = \frac{1}{\sqrt{N_\mathrm{T}}}, \hspace{5pt} \forall i,j, \nonumber \\
		& \mathbf{P}[m]\mathbf{P}^\textsf{H}[m] = \mathbf{I}_{N_\mathrm{S}},
		\end{align}
		where the unitary matrix $\mathbf{P}[m]\in \mathbb{C}^{K\times N_\mathrm{S}}$ is an auxiliary variable to provide a change of dimension between $\mathbf{F}_\mathrm{RF}\mathbf{F}_\mathrm{BB}[m]$ and $\mathbf{F}_\mathrm{R}$, which have different dimensions (i.e., $N_\mathrm{T}\times N_\mathrm{S}$ and $N_\mathrm{T}\times K$, respectively), without causing any distortion in the radar beampattern and $\mathbf{P}[m]\mathbf{P}^\textsf{H}[m] = \mathbf{I}_{K}$~\cite{radarCommLiuICASSP2019}.
	}

	\subsection{Problem Formulation}
	The aim of this work is designing the hybrid beamformer $\mathbf{F}_\mathrm{RF}\mathbf{F}_\mathrm{BB}[m]$ to simultaneously maximize  the spectral efficiency of the communications link and provide as much SNR as possible towards the radar targets by forming the beampattern of the transmit antenna array. To jointly solve the problems in (\ref{problemCom1}) and (\ref{problemRadar1}), we formulate the following problem, {\color{black}
		\begin{align}
		\label{problem1}
		&\min_{\mathbf{F}_\mathrm{RF},\{\mathbf{F}_\mathrm{BB}[m],\mathbf{P}[m]\}_{m\in \mathcal{M}}} \frac{1}{M}\sum_{m \in \mathcal{M}} \bigg(\eta \|\mathbf{F}_\mathrm{RF}\mathbf{F}_\mathrm{BB}[m]  - \hspace{-2pt} \mathbf{F}_\mathrm{C}[m]\|_\mathcal{F} \hspace{-2pt} \nonumber \\
		& \hspace{15pt}+ \bar{\eta}\|\mathbf{F}_\mathrm{RF}\mathbf{F}_\mathrm{BB}[m]  -  \mathbf{F}_\mathrm{R}\mathbf{P}[m]\|_\mathcal{F} \bigg) \nonumber \\
		&\subjectto \hspace{0pt} \sum_{m \in \mathcal{M}}\| \mathbf{F}_\mathrm{RF}\mathbf{F}_\mathrm{BB}[m] \|_\mathcal{F} = MN_\mathrm{S}, \nonumber \\
		&\hspace{30pt} |[\mathbf{F}_\mathrm{RF}]_{i,j}| = \frac{1}{\sqrt{N_\mathrm{T}}}, \hspace{5pt} \forall i,j\in \mathcal{S}, \nonumber \\
		&\hspace{30pt} |[\mathbf{F}_\mathrm{RF}]_{i,j}| = 0, \hspace{5pt} \forall i,j\in \bar{\mathcal{S}}, \nonumber \\
		& \hspace{30pt}\mathbf{P}[m]\mathbf{P}^\textsf{H}[m] = \mathbf{I}_{N_\mathrm{S}},
		\end{align}}
	where  $\mathcal{S}$ and $\bar{\mathcal{S}}$ denotes the set of non-zero and zero terms in $\mathbf{F}_\mathrm{RF}$ due to overlapped structure in (\ref{Frf}), respectively. In (\ref{problem1}), $0\leq \eta \leq 1$ provides the trade-off between the radar and communications tasks and $\bar{\eta} = (1 - \eta)$. If $\eta = 1$ ($\eta = 0$), (\ref{problem1}) corresponds to communications-only (radar-only) beamformer design problem. \textcolor{black}{The optimization problem (\ref{problem1}) is difficult solve because of several matrix variables $\mathbf{F}_\mathrm{RF}, \mathbf{F}_\mathrm{BB}[m],\mathbf{P}[m]$, and non-convex constraints. A common approach is to use alternating techniques, i.e., estimating the unknown variables one-by-one while fixing the others. While this approach does not guarantee the optimality, its convergence is proved in the relevant literature, e.g.,~\cite{hybridBFAltMin,elbirQuantizedCNN2019,elbir2020withoutCSI,radarCommLiuICASSP2019}.  }

	{\color{black}
		\textit{Assumption 1:} We assume that the THz channel matrix $\mathbf{H}[m]$ is available for CSI-based beamformer design. If necessary, the estimation of $\mathbf{H}[m]\in \mathbb{C}^{N_\mathrm{R}\times N_\mathrm{T}}$ can be performed via the following techniques~\cite{channelEstThz,channelEstThz2,ummimoTareq}, in which $N_\mathrm{R}\times N_\mathrm{T}$ array data can be used to construct the THz channel matrix in (\ref{channelMatrixCSI}). In addition, there exist model-free channel estimation techniques, e.g.,~\cite{elbir2019online,elbir2020_FL_CE}.
		
		\textit{Assumption 2:} We assume that the channel covariance matrix is available at the TX for statistical beamformer design (see, e.g., Section~\ref{sec:HBdesignCCM}).  In practice, the channel covariance matrix is estimated by several algorithms such as temporal averaging techniques and power angular spectrum estimation~\cite{covarianceCCMChannelEst} as well as model-based approaches, e.g.,~\cite{ccmEst_DL}.
	}

	\section{Model-Based Hybrid Beamformer Design}
	\label{sec:modelBased}
	In this part, we introduce our model-based hybrid beamformer design techniques relying on the CSI and channel covariance matrix-based channel information.
	
	\subsection{Hybrid Beamformer Design With CSI}
	\label{sec:HBdesignCSI} {\color{black}
		Denote $f(\mathbf{F}_\mathrm{RF}, \mathbf{F}_\mathrm{BB}[m],\mathbf{P}[m])$ as the cost function in (\ref{problem1}), which is rewritten as
		\begin{align}
		&f(\mathbf{F}_\mathrm{RF}, \mathbf{F}_\mathrm{BB}[m],\mathbf{P}[m]) = \|\eta[\mathbf{F}_\mathrm{RF}\mathbf{F}_\mathrm{BB}[m]  -  \mathbf{F}_\mathrm{C}[m]]\|_\mathcal{F} \nonumber \\
		& + \|\bar{\eta}[\mathbf{F}_\mathrm{RF}\mathbf{F}_\mathrm{BB}[m]  -  \mathbf{F}_\mathrm{R}\mathbf{P}[m]]\|_\mathcal{F}.
		\end{align}
		Then, using triangle inequality, we get 
		\begin{align}
		\label{costTriangleInequality}
		&f(\mathbf{F}_\mathrm{RF}, \mathbf{F}_\mathrm{BB}[m],\mathbf{P}[m]) \geq  \nonumber \\
		& \|\eta\mathbf{F}_\mathrm{RF}\mathbf{F}_\mathrm{BB}[m] -\eta\mathbf{F}_\mathrm{C}[m] + \bar{\eta} \mathbf{F}_\mathrm{RF}\mathbf{F}_\mathrm{BB}[m] -\bar{\eta} \mathbf{F}_\mathrm{R}\mathbf{P}[m] \|_\mathcal{F} \nonumber \\
		& =  \|\mathbf{F}_\mathrm{RF}\mathbf{F}_\mathrm{BB}[m] -\eta\mathbf{F}_\mathrm{C}[m]  - \bar{\eta} \mathbf{F}_\mathrm{R}\mathbf{P}[m] \|_\mathcal{F}.
		\end{align}
		Define $\mathbf{F}_\mathrm{CR}[m]\in \mathbb{C}^{N_\mathrm{T}\times N_\mathrm{S}}$ as the JRC beamformer as
		\begin{align}
		\mathbf{F}_\mathrm{CR}[m] = \eta\mathbf{F}_\mathrm{C} [m] + \bar{\eta} \mathbf{F}_\mathrm{R}\mathbf{P}[m],
		\end{align}
		and define the new cost function $\tilde{f}(\mathbf{F}_\mathrm{RF}, \mathbf{F}_\mathrm{BB}[m],\mathbf{P}[m]) $ as 
		\begin{align}
		\tilde{f}(\mathbf{F}_\mathrm{RF}, \mathbf{F}_\mathrm{BB}[m],\mathbf{P}[m]) \hspace{-3pt}= \hspace{-3pt} \|\mathbf{F}_\mathrm{RF}\mathbf{F}_\mathrm{BB}[m] -\mathbf{F}_\mathrm{CR}[m] \|_\mathcal{F},
		\end{align}
		where we have $\tilde{f}(\mathbf{F}_\mathrm{RF}, \mathbf{F}_\mathrm{BB}[m],\mathbf{P}[m])  \leq {f}(\mathbf{F}_\mathrm{RF}, \mathbf{F}_\mathrm{BB}[m],\mathbf{P}[m]) $ due to (\ref{costTriangleInequality}). Then, we rewrite the optimization problem (\ref{problem1}) as
		\begin{align}
		\label{problem2}
		&\min_{\mathbf{F}_\mathrm{RF},\{\mathbf{F}_\mathrm{BB}[m],\mathbf{P}[m]\}_{m \in \mathcal{M}}} \frac{1}{M}\sum_{m \in \mathcal{M}}\|\mathbf{F}_\mathrm{RF}\mathbf{F}_\mathrm{BB}[m] -\mathbf{F}_\mathrm{CR}[m] \|_\mathcal{F}  \nonumber \\
		&\subjectto \hspace{0pt}\sum_{m \in \mathcal{M}} \| \mathbf{F}_\mathrm{RF}\mathbf{F}_\mathrm{BB}[m] \|_\mathcal{F} = MN_\mathrm{S}, \nonumber \\
		&\hspace{30pt} |[\mathbf{F}_\mathrm{RF}]_{i,j}| = \frac{1}{\sqrt{N_\mathrm{T}}}, \hspace{5pt} \forall i,j\in \mathcal{S},  \nonumber \\
		&\hspace{30pt} |[\mathbf{F}_\mathrm{RF}]_{i,j}| = 0, \hspace{5pt} \forall i,j\in \bar{\mathcal{S}}, \nonumber\\
		& \hspace{30pt}\mathbf{P}[m]\mathbf{P}^\textsf{H}[m] = \mathbf{I}_{N_\mathrm{S}}.
		\end{align}
		
		The optimization problem in (\ref{problem2})  can be written in a compact form as
		\begin{subequations}
			\label{problem2OFDM}
			\begin{align}
			&\min_{\mathbf{F}_\mathrm{RF},\widetilde{\mathbf{F}}_\mathrm{BB},\widetilde{\mathbf{P}}} \hspace{3pt} \|\mathbf{F}_\mathrm{RF}\widetilde{\mathbf{F}}_\mathrm{BB} -\widetilde{\mathbf{F}}_\mathrm{CR} \|_\mathcal{F}  \nonumber \\
			&\subjectto \hspace{0pt} \| \mathbf{F}_\mathrm{RF}\widetilde{\mathbf{F}}_\mathrm{BB} \|_\mathcal{F} = MN_\mathrm{S},  \\
			&\hspace{30pt} |[\mathbf{F}_\mathrm{RF}]_{i,j}| = \frac{1}{\sqrt{N_\mathrm{T}}}, \hspace{5pt} \forall i,j\in \mathcal{S},   \label{FrfzeroConstraint1} \\
			&\hspace{30pt} |[\mathbf{F}_\mathrm{RF}]_{i,j}| = 0, \hspace{5pt} \forall i,j\in \bar{\mathcal{S}}, \label{FrfzeroConstraint2} \\
			& \hspace{30pt}\widetilde{\mathbf{P}}\widetilde{\mathbf{P}}^\textsf{H} = \mathbf{I}_{MN_\mathrm{S}},
			\end{align}
		\end{subequations}
		where $\widetilde{\mathbf{P}} = [\mathbf{P}[1],\cdots, \mathbf{P}[M]] $ is a $K\times MN_\mathrm{S}$ matrix, and we have  $	\widetilde{\mathbf{F}}_\mathrm{CR} = [ \mathbf{F}_\mathrm{CR}[1],
		\mathbf{F}_\mathrm{CR}[2], \cdots, \mathbf{F}_\mathrm{CR}[M] ]\in \mathbb{C}^{N_\mathrm{T}\times MN_\mathrm{S}}$
		and $	\widetilde{\mathbf{F}}_\mathrm{BB} = [ \mathbf{{F}}_\mathrm{BB}[1],  \mathbf{{F}}_\mathrm{BB}[2],  \cdots,  \mathbf{{F}}_\mathrm{BB}[M] ]$
		containing the beamformers for all subcarriers. $ \widetilde{\mathbf{F}}_\mathrm{CR}$ corresponds to the unconstrained radar-communications beamformer as $ \widetilde{\mathbf{F}}_\mathrm{CR} = \eta\widetilde{\mathbf{F}}_\mathrm{C}  + \bar{\eta} \mathbf{F}_\mathrm{R}\widetilde{\mathbf{P}}$, where $\widetilde{\mathbf{F}}_\mathrm{C} = [\mathbf{F}_\mathrm{C}[1],\cdots, \mathbf{F}_\mathrm{C}[M]]\in \mathbb{C}^{N_\mathrm{T}\times MN_\mathrm{S}}$.

		Now, the problem (\ref{problem2OFDM}) looks similar to the communications-only problem in (\ref{problemCom1}), and is solved via alternating minimization techniques suggested to solve (\ref{problemCom1}), e.g.,~\cite{heath2016overview,hybridBFAltMin}. In this case, $\mathbf{F}_\mathrm{RF}$, $\widetilde{\mathbf{F}}_\mathrm{BB}$ and $\widetilde{\mathbf{P}}$ are estimated one-by-one while the others are fixed. By fixing $\mathbf{F}_\mathrm{RF}$ and $\widetilde{\mathbf{F}}_\mathrm{BB}$,  $\widetilde{\mathbf{P}}$ is found via the SVD of the matrix $\mathbf{F}_\mathrm{R}^\textsf{H}\mathbf{F}_\mathrm{RF}\widetilde{\mathbf{F}}_\mathrm{BB}$, and for the $m$th subcarrier, we have 
		\begin{align}
		\label{computeP}
		\mathbf{P}[m] = \widetilde{\mathbf{U}}[m] \mathbf{I}_{K\times N_\mathrm{S}}\widetilde{\mathbf{V}}[m],
		\end{align}
		for which $\widetilde{\mathbf{U}}[m] \boldsymbol{\Sigma}[m]\widetilde{\mathbf{V}}^\textsf{H}[m] = \mathbf{F}_\mathrm{R}^\textsf{H}\mathbf{F}_\mathrm{RF}\mathbf{F}_\mathrm{BB}[m]$ and $\mathbf{I}_{K\times N_\mathrm{S}} = [\mathbf{I}_{K}, \hspace{1pt} \boldsymbol{0}_{K \times (N_\mathrm{S} -K)}]$~\cite{heath2016overview}. Similarly, when $\mathbf{F}_\mathrm{RF}$ and $\widetilde{\mathbf{P}}$ are fixed, $\widetilde{\mathbf{F}}_\mathrm{BB}$ is calculated. For the $m$th subcarrier,  $	\mathbf{F}_\mathrm{BB}[m] = \mathbf{F}_\mathrm{RF}^\dagger \mathbf{F}_\mathrm{CR}[m],$
		and it is normalized as $\mathbf{F}_\mathrm{BB}[m] = \frac{\sqrt{N_\mathrm{S}}}{\| \mathbf{F}_\mathrm{RF}\mathbf{F}_\mathrm{BB}[m]\|_\mathcal{F}}\mathbf{F}_\mathrm{BB}[m]$. 
	}
	The main challenge in (\ref{problem2OFDM}) is the estimation of $\mathbf{F}_\mathrm{RF}$ due to unit-modulus constraints. In fully-connected case, $\mathbf{F}_\mathrm{RF}$ is found via MO-based techniques and the optimal solution is readily obtained for the partially-connected structure via phase-rotation~\cite{hybridBFAltMin,radarCommLiuICASSP2019}. However, the design of $\mathbf{F}_\mathrm{RF}$ for the overlapped case is not straightforward due to the constraint (\ref{FrfzeroConstraint2}). Thus, we propose a MMO-based solution to account for (\ref{FrfzeroConstraint2}) in the following. 
	
	{\color{black}Assume that $\widetilde{\mathbf{F}}_\mathrm{BB}$ and $\widetilde{\mathbf{P}}$ are fixed, then  (\ref{problem2OFDM}) is written in vectorized form as
		\begin{align}
		\label{frf}
		&\min_{\mathbf{f}_\mathrm{RF}}  \|\mathbf{G}\mathbf{f}_\mathrm{RF} -\mathbf{f}_\mathrm{CR} \|_\mathcal{F}  \nonumber \\
		&\subjectto \hspace{3pt} |[\mathbf{f}_\mathrm{RF}]_{i}| = \frac{1}{\sqrt{N_\mathrm{T}}}, \hspace{5pt} \forall i \in \mathcal{V}, \nonumber \\
		&\hspace{30pt} |[\mathbf{f}_\mathrm{RF}]_{i}| = 0, \hspace{5pt} \forall i\in \bar{\mathcal{V}}, 
		\end{align}
		where $\mathbf{G} = (\widetilde{\mathbf{F}}_\mathrm{BB}^\textsf{T})\otimes \mathbf{I}_{N_\mathrm{T}} \in \mathbb{C}^{N_\mathrm{T}MN_\mathrm{S}\times N_\mathrm{T}N_\mathrm{RF}}$,  $\mathbf{f}_\mathrm{RF} = \mathrm{vec}\{\mathbf{F}_\mathrm{RF}\}\in \mathbb{C}^{N_\mathrm{T}N_\mathrm{RF}}$  and $\mathbf{f}_\mathrm{CR} = \mathrm{vec}\{\widetilde{\mathbf{F}}_\mathrm{CR}\}\in \mathbb{C}^{N_\mathrm{T}MN_\mathrm{S}}$. $\mathcal{V}$ and $\bar{\mathcal{V}}$ denote the set of non-zero and zero terms in $\mathbf{f}_\mathrm{RF}$, respectively. The sizes of $\mathcal{V}$ and $\bar{\mathcal{V}}$ depend on the selection of $\bar{M}$. As an example, for $N_\mathrm{T} = 100$, $N_\mathrm{RF} = 10$ and  $\bar{M} = N_\mathrm{RF}(N_\mathrm{T} - N_\mathrm{RF}+1)$, we have  $|\mathcal{V}| = 910$ and $|\bar{\mathcal{V}}| = 90$. Now, the aim is to exclude the portion of $\mathbf{G}$ and $\mathbf{f}_\mathrm{RF}$ corresponding to $\bar{\mathcal{V}}$ and find the portion of $\mathbf{f}_\mathrm{RF}$ corresponding to ${\mathcal{V}}$ so that we employ manifold optimization accordingly and all the elements of the unknown vector will obey unit-modulus constraints. }
	
	\begin{lemma} 
		\label{lem:sol1}
		Define  $\mathbf{G}_\mathcal{V}\in \mathbb{C}^{N_\mathrm{T}MN_\mathrm{S}\times T}$ and $\mathbf{f}_{\mathrm{RF}_\mathcal{V}}\in \mathbb{C}^{T}$ as the portion of $\mathbf{G}$ and $\mathbf{f}_\mathrm{RF}$ corresponding to ${\mathcal{V}}$, where $T$ is the number of remaining columns (entries) of $\mathbf{G}$ ($\mathbf{f}_\mathrm{RF}$) after excluding the terms related to $\bar{\mathcal{V}}$. Then, the optimization problem in (\ref{frf}) is equivalent to
		\begin{align}
		\label{frf2}
		&\min_{\mathbf{f}_{\mathrm{RF}_\mathcal{V}}}  \|\mathbf{G}_\mathcal{V}\mathbf{f}_{\mathrm{RF}_\mathcal{V}} -\mathbf{f}_\mathrm{CR} \|_\mathcal{F}  \nonumber \\
		&\subjectto \hspace{3pt} |[\mathbf{f}_{\mathrm{RF}_\mathcal{V}}]_{i}| = \frac{1}{\sqrt{N_\mathrm{T}}}, \hspace{5pt} \forall i \in \mathcal{V}.
		\end{align}
	\end{lemma}
	\begin{IEEEproof} The cost of problem (\ref{frf}) is
		\begin{align}
		\label{frfMod}
		\|[\mathbf{G}_\mathcal{V},  &\mathbf{G}_{\bar{\mathcal{V}}}]        \left[ \begin{array}{c}
		\mathbf{f}_{\mathrm{RF}_\mathcal{V}}\\
		\mathbf{f}_{\mathrm{RF}_{\bar{\mathcal{V}}}}
		\end{array}\right] - 
		\mathbf{f}_\mathrm{CR}
		\|_\mathcal{F} \nonumber \\
		&=  \|
		\mathbf{G}_\mathcal{V}\mathbf{f}_{\mathrm{RF}_\mathcal{V}} +
		\mathbf{G}_{\bar{\mathcal{V}}}\mathbf{f}_{\mathrm{RF}_{\bar{\mathcal{V}}}}  - \mathbf{f}_\mathrm{CR} \|_\mathcal{F}\nonumber\\
		& =  \|
		\mathbf{G}_\mathcal{V}\mathbf{f}_{\mathrm{RF}_\mathcal{V}} -\mathbf{f}_\mathrm{CR}   \|_\mathcal{F}.
		\end{align}
		Since  $\mathbf{f}_{\mathrm{RF}_{\bar{\mathcal{V}}}}= \mathbf{0}_{N_\mathrm{T}N_\mathrm{RF}-T\times 1}$, we have $\mathbf{G}_{\bar{\mathcal{V}}} \mathbf{f}_{\mathrm{RF}_{\bar{\mathcal{V}}}} = \mathbf{0}_{N_\mathrm{T}MN_\mathrm{S}\times (N_\mathrm{T}N_\mathrm{RF} - T)}$, and thus (\ref{frf}) is equivalent to (\ref{frf2}). 
	\end{IEEEproof}
	
	Using Lemma~\ref{lem:sol1}, we solve (\ref{frf2}) via MO, for which the search space is regarded as a \textit{Riemannian submanifold} $\mathcal{M}$ of complex plane $\mathbb{C}^{T}$ since $\mathbf{f}_{\mathrm{RF}_\mathcal{V}}\in \mathbb{C}^{T}$ forms a complex circle manifold, i.e., $\mathcal{M}_{cc}^{T} = \{ \mathbf{f}_{\mathrm{RF}_\mathcal{V}}\in \mathbb{C}^{T} : |[\mathbf{f}_{\mathrm{RF}_\mathcal{V}}]_1| = |[\mathbf{f}_{\mathrm{RF}_\mathcal{V}}]_2|= \dots = |[\mathbf{f}_{\mathrm{RF}_\mathcal{V}}]_T| = \frac{1}{\sqrt{N_\mathrm{T}}}\} $ due to unit-modulus constraint. Define the Riemannian gradient at $\mathbf{f}_{\mathrm{RF}_\mathcal{V}}$, $\mathrm{grad}f(\mathbf{f}_{\mathrm{RF}_\mathcal{V}})$ as the orthogonal projection of the 
	Euclidean gradient $\nabla f(\mathbf{f}_{\mathrm{RF}_\mathcal{V}})$ onto the tangent space of $\mathbf{f}_{\mathrm{RF}_\mathcal{V}}$, i.e.,
	\begin{align}
	\mathrm{grad}f(\mathbf{f}_{\mathrm{RF}_\mathcal{V}}) =  \nabla f(\mathbf{f}_{\mathrm{RF}_\mathcal{V}}) - \operatorname{Re}\{ \nabla f(\mathbf{f}_{\mathrm{RF}_\mathcal{V}}) \odot \mathbf{f}_{\mathrm{RF}_\mathcal{V}}^* \} \odot \mathbf{f}_{\mathrm{RF}_\mathcal{V}},\nonumber
	\end{align}
	where the Euclidean gradient of the cost function in (\ref{frf2}) is given by
	\begin{align}
	\label{eq:gradF}
	\nabla f(\mathbf{f}_{\mathrm{RF}_\mathcal{V}}) = -2\mathbf{G}_\mathcal{V}^\textsf{H} [ {\mathbf{f}}_\mathrm{CR}- \mathbf{G}_\mathcal{V} \mathbf{f}_{\mathrm{RF}_\mathcal{V}} ].
	\end{align}
	After defining the cost function and the gradient as in (\ref{frf2}) and (\ref{eq:gradF}), the remaining part of the solution is similar to the conventional manifold optimization algorithm~\cite{manopt}. This is done via the conjugate gradient descent technique iteratively such that $\mathbf{f}_{\mathrm{RF}_\mathcal{V}}^{(k+1)}$ at the $k$th iteration is obtained with the update rule
	\begin{align}
	\label{Frf_update}
	\mathbf{f}_{\mathrm{RF}_\mathcal{V}}^{(k+1)} = \frac{ (\mathbf{f}_{\mathrm{RF}_\mathcal{V}}^{(k)} + \alpha_k {\bf d }^{(k)}) }  {|(\mathbf{f}_{\mathrm{RF}_\mathcal{V}}^{(k)} + \alpha_k \mathbf{f}_{\mathrm{RF}_\mathcal{V}}^{(k)})|},
	\end{align}
	where $\alpha_{k}$ is Armijo backtracking line search step size~\cite{manopt} and  $\mathbf{d}_k$ denotes the direction of decrease, which is defined as $	{\bf d}_{k} =- \mathrm{grad}f(\mathbf{f}_{\mathrm{RF}_\mathcal{V}}^{(k)}) + \beta_k \bar{\bf d}^{(k-1)},$	for which $\mathrm{grad}f(\mathbf{f}_{\mathrm{RF}_\mathcal{V}}^{(k)})$ denotes the Riemannian gradient at the $k$th iteration and $\beta_k$ is the Polak-Ribiere parameter. $\bar{\bf d}^{(k)}$ is the vector transport of conjugate direction ${\bf d}^{(k)}$, which is defined as $	\bar{\bf d}^{(k)} = {\bf d}^{(k)} - \operatorname{Re}\{{\bf d}^{(k)} \odot \mathbf{f}_{\mathrm{RF}_\mathcal{V}}^{(k+1)} \} \odot\mathbf{f}_{\mathrm{RF}_\mathcal{V}}^{(k+1)},$
	where $\mathbf{f}_{\mathrm{RF}_\mathcal{V}}^{(k+1)}$ is directly obtained from (\ref{Frf_update}) and ${\bf d}_{0} =- \mathrm{grad}f(\mathbf{f}_{\mathrm{RF}_\mathcal{V}}^{(0)})$. The optimization process is initialized from a random point, i.e., $[\mathbf{f}_{\mathrm{RF}_\mathcal{V}}^{(0)}]_{t} = e^{j\bar{\theta}_{t}}$ where $\bar{\theta}_{t} \sim \mathrm{uniform}([0,2\pi))$, $t = 1,\dots,T$. Once the non-zero entries of $\mathbf{f}_\mathrm{RF}$, i.e., $\mathbf{f}_{\mathrm{RF}_\mathcal{V}}$ is optimized, then RF precoder $\mathbf{F}_\mathrm{RF}\in \mathbb{C}^{N_\mathrm{T}\times N_\mathrm{RF}}$ is reconstructed from $\mathbf{f}_{\mathrm{RF}_\mathcal{V}}$ according to the index sets $\mathcal{V}$, $\bar{\mathcal{V}}$, $\mathcal{S}$  and $\bar{\mathcal{S}}$.
	
	In (\ref{problem2OFDM}), the convergence to an optimum solution is guaranteed such that the Euclidean distance between the radar-communications beamformer $\widetilde{\mathbf{F}}_\mathrm{CR}$ and the hybrid beamformer $\mathbf{F}_\mathrm{RF}\widetilde{\mathbf{F}}_\mathrm{BB}$ is minimized \cite{hybridBFAltMin}. We present the algorithmic steps of the CSI-based hybrid beamformer design in Algorithm~\ref{alg:CSI}, \textcolor{black}{which also includes beam split correction procedure as described in Appendix~\ref{appBeamsplit}}. The implementation of the iterative algorithm takes no more than $10$ iterations while the MMO steps requires approximately $20$ sub-iterations for the settings $N_\mathrm{T} = 1024$, $Q = 9$ and $N_\mathrm{RF} = 10$. \textcolor{black}{Furthermore, the complexity order of the MMO algorithm is the same as the conventional manifold optimization algorithm~\cite{hybridBFAltMin}, and it is mainly due to the computation of the conjugate gradient in (\ref{eq:gradF}) which is $\mathcal{O}(N_\mathrm{iter}N_\mathrm{T}M N_\mathrm{S} T^2) $, where $N_\mathrm{iter}$ denotes the number of iterations and $T$ is the number of non-zero entries of $\mathbf{F}_\mathrm{RF}$~\cite{radarCommLiuICASSP2019,hybridBFAltMin,manopt}.  }

	\begin{algorithm}[t]
		{\color{black}
			\begin{algorithmic}[1]
				\caption{Hybrid beamforming for joint ultra-massive MIMO radar-communications}
				\Statex {\textbf{Input:} $\eta$, $\widetilde{\mathbf{F}}_\mathrm{C}$, $\mathbf{F}_\mathrm{R}$}.
				\label{alg:CSI}
				\Statex {\textbf{Output:} $\mathbf{F}_\mathrm{RF}$, ${\mathbf{F}}_\mathrm{BB}[m]$ and ${\mathbf{F}}_\mathrm{BB}^\mathrm{c}[m]$ for $m\in \mathcal{M}$.}
				\State Initialize with random $\mathbf{F}_\mathrm{RF}\in \mathbb{C}^{N_\mathrm{T}\times N_\mathrm{RF}}$, $\mathbf{F}_\mathrm{BB}[m]\in \mathbb{C}^{N_\mathrm{RF}\times N_\mathrm{S}}$ and $\mathbf{P}[m]\in \mathbb{C}^{K\times N_\mathrm{S}}$.
				\State $\widetilde{\mathbf{F}}_\mathrm{CR} = \eta\widetilde{\mathbf{F}}_\mathrm{C}  + \bar{\eta} \mathbf{F}_\mathrm{R}\widetilde{\mathbf{P}}$.
				\State Construct $\mathcal{S}$ and $\bar{\mathcal{S}}$ depending on the structure of $\mathbf{F}_\mathrm{RF}$.
				\State \textbf{while}
				\State \indent Construct $\widetilde{\mathbf{P}} = [\mathbf{P}[1],\dots, \mathbf{P}[M]]$, where \par \indent 
				$\mathbf{P}[m] = \widetilde{\mathbf{U}}[m] \mathbf{I}_{K\times N_\mathrm{S}}\widetilde{\mathbf{V}}[m]$ and  $\widetilde{\mathbf{U}}[m] \boldsymbol{\Sigma}[m]\widetilde{\mathbf{V}}^\textsf{H}[m] =$ \par \indent $ \mathbf{F}_\mathrm{R}^\textsf{H}\mathbf{F}_\mathrm{RF}\mathbf{F}_\mathrm{BB}[m]$ for $ m \in \mathcal{M}$.
				\State \indent Compute $\mathbf{F}_\mathrm{BB}[m]$ as $\mathbf{F}_\mathrm{BB}[m] = \mathbf{F}_\mathrm{RF}^\dagger \mathbf{F}_\mathrm{CR}[m]$ and  \par \indent normalize as  $\mathbf{F}_\mathrm{BB}[m] = \frac{\sqrt{N_\mathrm{S}}}{\| \mathbf{F}_\mathrm{RF}\mathbf{F}_\mathrm{BB}[m]\|_\mathcal{F}}\mathbf{F}_\mathrm{BB}[m]$.
				\State \indent Use  $\mathcal{S}$ and $\bar{\mathcal{S}}$ and find $\mathbf{F}_\mathrm{RF}$ with the MMO algorithm \par \indent in (\ref{problem2}) and (\ref{frf}).
				\State \textbf{until} convergence
				\State Beam split correction: $	{\mathbf{F}}_\mathrm{BB}^\mathrm{c}[m] = \big({\mathbf{F}}_\mathrm{RF}\big)^\dagger  {\mathbf{F}}_\mathrm{RF}^\mathrm{c}[m] {\mathbf{F}}_\mathrm{BB}[m]$ for $m\in \mathcal{M}$ as in (\ref{beamSplitEquation}).
			\end{algorithmic}
		}
	\end{algorithm}

	The hybrid beamformer design problem in (\ref{problem2OFDM}) should be solved for every $\eta$. Furthermore,  $\mathbf{P}[m]$ changes even if $\mathbf{F}_\mathrm{R}$ is kept fixed since it matches the hybrid beamformer to the unconstrained radar-only beamformer. As a special case, where $\mathbf{P}[m] = \mathbf{I}_\mathrm{K}$, i.e., $K= N_\mathrm{S}$, the following lemma shows that the solution of (\ref{problem2OFDM}) is obtained from the linear combination of the solutions of communications- and radar-only problems in (\ref{problemCom1}) and (\ref{problemRadar1}), respectively.
	
	{\color{black}
		\begin{lemma}
			\label{lem:sol2} Denote $\check{\mathbf{F}}_\mathrm{C}\in \mathbb{C}^{N_\mathrm{T}\times MN_\mathrm{S}}$ and $\check{\mathbf{F}}_\mathrm{R}\in \mathbb{C}^{N_\mathrm{T}\times MN_\mathrm{S}}$ as the hybrid beamforming solutions of  (\ref{problemCom1}) and (\ref{problemRadar1}), respectively. Then, the solution of (\ref{problem2OFDM}) is 
			\begin{align}
			\label{lemma2Claim}
			\check{\mathbf{F}}_\mathrm{CR} = \eta \check{\mathbf{F}}_\mathrm{C} + (1 - \eta)\check{\mathbf{F}}_\mathrm{R},
			\end{align}
			if $K= N_\mathrm{S}$ and $\mathbf{P}[m]=\mathbf{I}_K$ for $m\in \mathcal{M}$.
		\end{lemma}
		\begin{IEEEproof} If $\mathbf{P}= \mathbf{I}_\mathrm{K}$, we have $\widetilde{\mathbf{F}}_\mathrm{R} =[ \underbrace{\mathbf{F}_\mathrm{R},\cdots, \mathbf{F}_\mathrm{R}}_{M}]\in \mathbb{C}^{N_\mathrm{T}\times MK}$, and  the solution of (\ref{problem2OFDM}) involves the alternations only between $\mathbf{F}_\mathrm{RF}$ and $\widetilde{\mathbf{F}}_\mathrm{BB}\in \mathbb{C}^{N_\mathrm{RF}\times MN_\mathrm{S}}$. Then, rewrite the cost function of (\ref{problem2OFDM}) as
			\begin{align}
			\label{lemmaCost1}
			f(\check{\mathbf{F}}_\mathrm{C},\check{\mathbf{F}}_\mathrm{R}) =\eta \|\check{\mathbf{F}}_\mathrm{C} \hspace{-2pt} - \hspace{-2pt} \widetilde{\mathbf{F}}_\mathrm{C}\|_\mathcal{F}  +(1 - \eta)\|\check{\mathbf{F}}_\mathrm{R}\hspace{-2pt}  - \hspace{-2pt} \widetilde{\mathbf{F}}_\mathrm{R}\|_\mathcal{F}, 
			\end{align}
			from which it is clear that $\check{\mathbf{F}}_\mathrm{C}$ and $\check{\mathbf{F}}_\mathrm{R}$ correspond to the communications- and radar-only hybrid beamforming solutions in (\ref{problemCom1}) and (\ref{problemRadar1}) for $\eta = 1$ and $\eta = 0$, respectively. 
			Using the triangle inequality expression in (\ref{costTriangleInequality}), $f(\check{\mathbf{F}}_\mathrm{C},\check{\mathbf{F}}_\mathrm{R})$  in (\ref{lemmaCost1}) is lower-bounded by $\bar{f}(\check{\mathbf{F}}_\mathrm{C},\check{\mathbf{F}}_\mathrm{R})$ as 
			\begin{align}
			\underbrace{\|\eta\big[\check{\mathbf{F}}_\mathrm{C}  -  \mathbf{F}_\mathrm{C}\big] + (1 - \eta)\big[\check{\mathbf{F}}_\mathrm{R}  -  \widetilde{\mathbf{F}}_\mathrm{R}\big]\|_\mathcal{F}}_{\overset{\Delta}{=}\bar{f}(\check{\mathbf{F}}_\mathrm{C},\check{\mathbf{F}}_\mathrm{R})}  \leq  {f}(\check{\mathbf{F}}_\mathrm{C},\check{\mathbf{F}}_\mathrm{R}),
			\end{align}
			where $\bar{f}(\check{\mathbf{F}}_\mathrm{C},\check{\mathbf{F}}_\mathrm{R})$ can be rewritten as 
			\begin{align}
			\bar{f}(\check{\mathbf{F}}_\mathrm{C},\check{\mathbf{F}}_\mathrm{R})
			& = \|\eta\check{\mathbf{F}}_\mathrm{C}  + (1 - \eta)\check{\mathbf{F}}_\mathrm{R} - \eta\mathbf{F}_\mathrm{C}  - \bar{\eta}\widetilde{\mathbf{F}}_\mathrm{R}\|_\mathcal{F} \nonumber\\
			& = \| \check{\mathbf{F}}_\mathrm{CR} - (1 - \eta)\mathbf{F}_\mathrm{C}  - (1 - \eta)\widetilde{\mathbf{F}}_\mathrm{R} \|_\mathcal{F}\nonumber\\
			& = \| \check{\mathbf{F}}_\mathrm{CR} - \mathbf{F}_\mathrm{CR}  \|_\mathcal{F},
			\end{align}
			which gives (\ref{lemma2Claim}) as the linear combination of $\check{\mathbf{F}}_\mathrm{C}$ and $\check{\mathbf{F}}_\mathrm{R}$ by depending on $\eta$. 
		\end{IEEEproof}
		
		This analysis allows us to design the JRC hybrid beamformer as a function of $\eta$ after solving the communications- and radar-only problems (\ref{problemCom1}) and (\ref{problemRadar1}), respectively, instead of solving (\ref{problem2OFDM}) for every $\eta$.
		
	}
	\subsection{Hybrid Beamformer Design With Channel Covariance Matrix}
	\label{sec:HBdesignCCM}
	Instead of designing the hybrid beamformer $\mathbf{F}_\mathrm{RF}\mathbf{F}_\mathrm{BB}[m]$ with respect to $\mathbf{H}[m]$, the usage of the channel covariance matrix provides lower channel overhead via infrequent updates of the THz channel information between the RX and the TX. However, this approach has the cost of slight performance loss in the spectral efficiency due to long-term statistics of the channel information. The usage of channel covariance matrix is particularly helpful in THz transmission compared to the mm-Wave case due to smaller number of LoS/NLoS signal components, which reduces the angular spread of the received signals~\cite{ummimoTareqOverview}.  To exploit the structure of channel covariance matrix-based hybrid beamforming, we first introduce the channel covariance matrix model to derive the near-optimal unconstrained channel covariance matrix-based beamformer $\bar{\mathbf{F}}_\mathrm{C}[m]\in \mathbb{C}^{N_\mathrm{T}\times N_\mathrm{S}}$ via the eigendecomposition of the channel covariance matrix $\mathbf{C}[m]\in \mathbb{C}^{N_\mathrm{T}\times N_\mathrm{T}}$. Rewrite (\ref{channelMatrixCSI}) as
	{\color{black}
		\begin{align}
		\mathbf{H}[m] = \gamma \mathbf{B}_\mathrm{R}^m\boldsymbol{\Gamma} \mathbf{B}_\mathrm{T}^{m^\textsf{H}},
		\end{align}
		where $\mathbf{B}_\mathrm{R}^m = [\mathbf{A}_\mathrm{R}^m(\Theta_{1}),\dots, \mathbf{A}_\mathrm{R}^m(\Theta_{L})]\in \mathbb{C}^{N_\mathrm{R}\times QL}$ and $\mathbf{B}_\mathrm{T}^m = [\mathbf{A}_\mathrm{T}^m(\Psi_1),\dots, \mathbf{A}_\mathrm{T}^m(\Psi_L)]\in \mathbb{C}^{N_\mathrm{T}\times QL}$  steering matrices of $L$ paths, respectively. $\boldsymbol{\Gamma}^m = \mathrm{blkdiag}\{\alpha_{1,m} \mathbf{I}_Q,\dots,\alpha_{L,m} \mathbf{I}_Q  \}\in \mathbb{C}^{QL\times QL}$ is a diagonal matrix which includes the path gains. Using the property that the channel gains are independent random variables, we write the covariance of the channel at the TX, i.e.,  $\mathbf{C}[m] = \mathbb{E}\{\mathbf{H}^\textsf{H}[m] \mathbf{H}[m]  \}$ as
		\begin{align}
		\label{Cov1}
		\mathbf{C}[m] = \gamma^2 \mathbb{E}_\mathbf{H} \{ \mathbf{B}_\mathrm{T}^m \boldsymbol{\Gamma}^{m^\textsf{H}} \mathbf{B}_\mathrm{R}^{m^\textsf{H}} \mathbf{B}_\mathrm{R}^m \boldsymbol{\Gamma}^m \mathbf{B}_\mathrm{T}^{m^\textsf{H}}  \},
		\end{align}
		where the expectation is performed over $\mathbf{H}[m]$. Taking statistical expectation over the AoA/AoD angles and the channel gains, respectively, (\ref{Cov1}) becomes
		\begin{align}
		\mathbf{C}[m] = \gamma^2 \mathbb{E}_{\Psi} \{ \mathbf{B}_\mathrm{T}^m \mathbb{E}_{\alpha}\{    \boldsymbol{\Gamma}^{m^\textsf{H}} \mathbb{E}_{\Theta} \{\mathbf{B}_\mathrm{R}^{m^\textsf{H}} \mathbf{B}_\mathrm{R}^m \} \boldsymbol{\Gamma}^m \} \mathbf{B}_\mathrm{T}^{m^\textsf{H}}  \}, 
		\end{align}
		for which we have $\mathbb{E}_{\Theta} \{\mathbf{B}_\mathrm{R}^{m^\textsf{H}} \mathbf{B}_\mathrm{R}^m \} = \mathbf{I}_{QL}$ and $\mathbb{E}_{\alpha}\{    \boldsymbol{\Gamma}^{m^\textsf{H}}  \boldsymbol{\Gamma}^m \} = \widetilde{\boldsymbol{\Gamma}}^m= \mathrm{blkdiag}\{ \sigma_{\alpha_{1,m}}^2 \mathbf{I}_\mathrm{Q},\dots,\sigma_{\alpha_{L,m}}^2\mathbf{I}_\mathrm{Q}  \}$ due to the independent zero-mean channel gains. Thus, we finally get
		\begin{align}
		\label{covariance}
		\mathbf{C}[m] =& \gamma^2 \mathbf{B}_\mathrm{T}^m\widetilde{\boldsymbol{\Gamma}}^m\mathbf{B}_\mathrm{T}^{m^\textsf{H}} \nonumber \\
		&=\gamma^2 \sum_{l=1}^{L}\sigma_{\alpha_{l,m}}^2 \mathbb{E}\{\mathbf{A}_\mathrm{T}^m(\Psi_l) \mathbf{A}_\mathrm{T}^{m^\textsf{H}}(\Psi_l)  \}.
		\end{align}
		Since the path gains of the NLoS paths (i.e., $l = 2,\dots, L$) are significantly smaller as compared to the LoS path in THz channels, the channel covariance matrix in (\ref{covariance}) can be approximated as
		\begin{align}
		\label{covarianceApp}
		\mathbf{C}[m] \approx \gamma^2 \sigma_{\alpha_{1,m}}^2 \mathbf{A}_\mathrm{T}^m(\Psi_{1})\mathbf{A}_\mathrm{T}^m(\Psi_{1})^\textsf{H}.
		\end{align}
		Compared to the CSI in (\ref{channelMatrixCSI}), the channel covariance matrix in  (\ref{covarianceApp}) only preserves the channel statistics, such as variance of received path gains at the TX $\sigma_{\alpha_{1,m}}^2$.	While the channel covariance matrix does not provide us the complete instantaneous channel knowledge as of $\mathbf{H}[m]$, it has lower channel feedback since the RX only needs to send $\sigma_{\alpha_{1,m}}^2$  and the mean AoD angle of the LoS path $\Psi_{1}$.
		
		Using the channel covariance matrix in (\ref{covarianceApp}), the optimal unconstrained statistical beamformer  $\bar{\mathbf{F}}_\mathrm{C}[m]\in \mathbb{C}^{N_\mathrm{T}\times N_\mathrm{S}}$ is designed via the following quadratic problem, i.e.,
		\begin{align}
		\label{FoptStat}
		\bar{\mathbf{F}}_\mathrm{C}[m] = &\arg \maximize_{\tilde{\mathbf{F}}} \hspace{2pt} ||\tilde{\mathbf{F}}^\textsf{H} [m]\mathbf{C}[m] \tilde{\mathbf{F}}[m]||_\mathcal{F} \nonumber \\
		&\subjectto || \tilde{\mathbf{F}}[m] ||_\mathcal{F} = N_\mathrm{S},
		\end{align}
		whose solution is readily found as the linear combination of the eigenvectors corresponding to the $N_\mathrm{S}$ largest eigenvalues~\cite{statisticalChannelModel1}.

		Once the unconstrained statistical beamformer is obtained, the next task is to determine the analog precoder $\mathbf{F}_\mathrm{RF}$ via the following optimization problem, i.e.,
		\begin{align}
		\label{problem3}
		&\underset{\mathbf{F}_\mathrm{RF},\widetilde{\mathbf{F}}_\mathrm{BB},\widetilde{\mathbf{P}}}{\minimize}\hspace{3pt} \big|\big|   \mathbf{F}_\mathrm{RF}\widetilde{\mathbf{F}}_\mathrm{BB} - \widetilde{\bar{\mathbf{F}}}_\mathrm{CR} \big|\big|_\mathcal{F}
		\nonumber \\
		&\subjectto \; |[\mathbf{F}_\mathrm{RF}]_{i,j} |=\frac{1}{\sqrt{N_\mathrm{T}}},  \nonumber \\
		&\hspace{30pt}\| \mathbf{{F}}_\mathrm{RF}\widetilde{\mathbf{F}}_\mathrm{BB}\|_{\mathcal{F}}  =  MN_\mathrm{S}, \nonumber\\
		& \hspace{30pt}\widetilde{\mathbf{P}}\widetilde{\mathbf{P}}^\textsf{H} = \mathbf{I}_{MN_\mathrm{S}},
		\end{align}
		where $\widetilde{\bar{\mathbf{F}}}_\mathrm{CR} = [\bar{\mathbf{F}}_\mathrm{CR}[1],\cdots, \bar{\mathbf{F}}_\mathrm{CR}[M]]\in \mathbb{C}^{N_\mathrm{T}\times MN_\mathrm{S}}$. The channel covariance matrix-based problem in (\ref{problem3}) is similar to (\ref{problem2OFDM}) and the solution is obtained by following the similar procedure presented in Algorithm~\ref{alg:CSI} by replacing $\widetilde{\mathbf{F}}_\mathrm{C}$ with $\widetilde{\bar{\mathbf{F}}}_\mathrm{C}$.}

	\begin{figure}[t]
		\centering
		{\includegraphics[draft=false,width=\columnwidth]{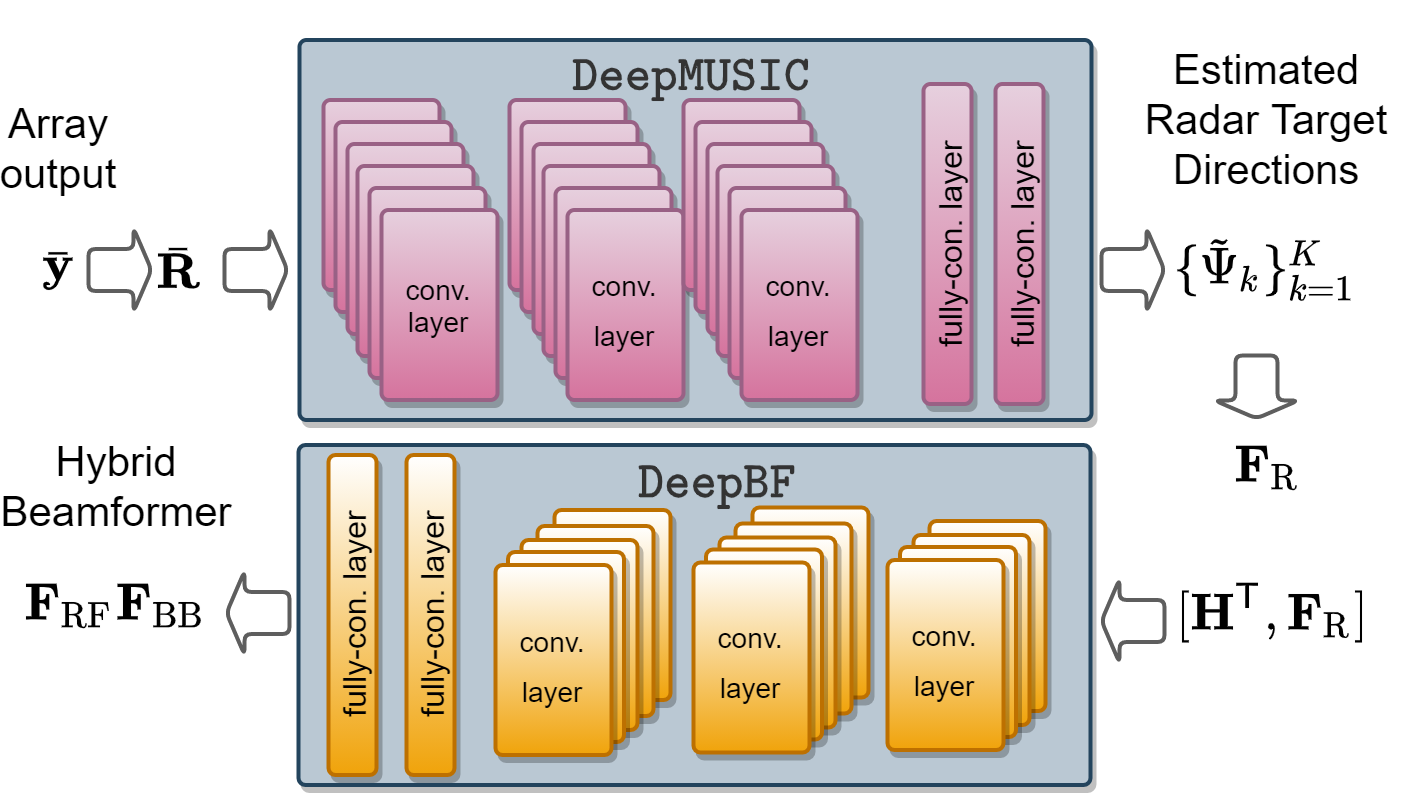} } 
		\caption{Model-free hybrid beamforming framework, in which \texttt{DeepMUSIC}~\cite{elbir_DL_MUSIC} and \texttt{DeepBF} are employed to predict radar target directions and hybrid beamformer weights, respectively.
		}
		\label{fig_Model}
	\end{figure}

	\section{Model-Free Hybrid Beamforming}
	\label{sec:modelFree}
	In this section, we introduce our model-free approach for hybrid beamforming by designing a CNN model. While the design of the learning model is relatively easier for  communications-only~\cite{elbirQuantizedCNN2019,elbir2019online} and radar-only~\cite{elbir_DL_MUSIC} problems, the joint scenario involves several matrix variables, such as \textcolor{black}{$\mathbf{H}[m]$, $\mathbf{F}_\mathrm{C}[m]$, $\mathbf{F}_\mathrm{RF}$, $\mathbf{F}_\mathrm{BB}[m]$, $\mathbf{F}_\mathrm{R}$ }and $\mathbf{P}[m]$, which make the problem very challenging. Another difficulty is due to the size of these variables, which are huge because of the large number of antennas deployed in THz scenario. \textcolor{black}{The usage of GoSA structure reduces the size of the beamformer data by the order of $Q$. Furthermore, quantized learning models further reduce the computational and memory complexity of the training process for large datasets~\cite{elbirQuantizedCNN2019}.}
	
	To efficiently train the model while maintaining satisfactory learning performance, we first adopt the \texttt{DeepMUSIC} model of~\cite{elbir_DL_MUSIC} to estimate the radar target directions and construct $\mathbf{F}_\mathrm{R}$, as shown in Fig.~\ref{fig_Model}. \textcolor{black}{Furthermore, we assume the narrowband model for simplicity while the design of the learning model can be done for wideband scenario by processing the input data for all subcarriers as shown in~\cite{elbir2019online}.} To this end, the array output at the TX is utilized.  The radar collects the reflected signal from the targets as
	\begin{align}
	\bar{\mathbf{y}}(t_i) = \sum_{k = 1}^{K}\mathbf{W}_\mathrm{RF}{\mathbf{a}}_\mathrm{T}(\Phi_k){\mathbf{a}}_\mathrm{T}^\textsf{T}(\Phi_k) {r}_k(t_i) + \mathbf{W}_\mathrm{RF}\bar{\mathbf{n}}(t_i),
	\end{align}
	where $t_i$ denotes the sample index for $i = 1,\dots, T_R$, where $T_R$ is the number of snapshots, $r_k(t_i)$ represents the reflection coefficient of the transmitted  signal corresponding to the $k$th target, and $\bar{\mathbf{n}}(t_i)$ denotes the $N_\mathrm{T}\times 1$ noise term. $\mathbf{W}_\mathrm{RF}{\mathbf{a}}_\mathrm{T}(\Phi_k)$ denotes the actual steering vector after processing via the analog combiner $\mathbf{W}_\mathrm{RF}\in \mathbb{C}^{N_\mathrm{RF}\times N_\mathrm{T}}$~\cite{radarCommSurvey}. Then, the sample covariance matrix is computed as $\bar{\mathbf{R}} = \frac{1}{T_R} \sum_{i = 1}^{T_R} \bar{\mathbf{y}}(t_i)\bar{\mathbf{y}}^\textsf{H}(t_i) $. $\bar{\mathbf{R}}$ is input to \texttt{DeepMUSIC} to obtain the MUSIC spectra at the output~\cite{elbir_DL_MUSIC}. After performing peak-finding on the resultant spectra, the estimated target locations are acquired and the corresponding $\mathbf{F}_\mathrm{RF}$ are constructed.
	
	To obtain the hybrid beamformers, we design another model, which is called \texttt{DeepBF} (Fig.~\ref{fig_Model}), for which $\mathbf{F}_\mathrm{R}$ is utilized together with the channel matrix $\mathbf{H}$   to represent the inputs for  radar- and communications-only tasks, respectively. The input of \texttt{DeepBF} is then $\boldsymbol{\Xi} = [\mathbf{H}^\textsf{T},\mathbf{F}_\mathrm{R}]\in \mathbb{C}^{N_\mathrm{T}\times (N_\mathrm{R} +K)}$. If the partially-connected array is assumed, then $\mathbf{F}_\mathrm{R}$ is squeezed into a vector yielding an ${N_\mathrm{T}\times (N_\mathrm{R} +1)}$ input. The ``channels" of \texttt{DeepBF} are designed as the real, imaginary values of $\boldsymbol{\Xi}$, the input size of \texttt{DeepBF} is ${N_\mathrm{T}\times (N_\mathrm{R} +K)\times 2}$ for a single data sample. The output of \texttt{DeepBF} is designed as the real and imaginary values of the hybrid beamformer  $\mathbf{F}_\mathrm{RF}\mathbf{F}_\mathrm{BB}$, i.e., $\boldsymbol{\xi} = [\mathrm{vec}\{\operatorname{Re}\{\mathbf{F}_\mathrm{RF}\mathbf{F}_\mathrm{BB}\}\}^\textsf{T}, \mathrm{vec}\{\operatorname{Im}\{\mathbf{F}_\mathrm{RF}\mathbf{F}_\mathrm{BB}\}\}^\textsf{T}]^\textsf{T}\in \mathbb{R}^{2N_\mathrm{T}N_\mathrm{S}}$. Thus, the learning model constructs the non-linear relationship $\mathcal{L} : \mathbb{R}^{{N_\mathrm{T}\times (N_\mathrm{R} +K)}} \rightarrow \mathbb{R}^{2N_\mathrm{T}N_\mathrm{S}} $ as $	\mathcal{L}(\boldsymbol{\Xi}, \boldsymbol{\theta}) = \boldsymbol{\xi},$	where $\boldsymbol{\theta}$ represents the learnable parameters of \texttt{DeepBF}.

	\begin{figure}[t]
		\centering
		\subfloat[] {\includegraphics[draft=false,width=\columnwidth]{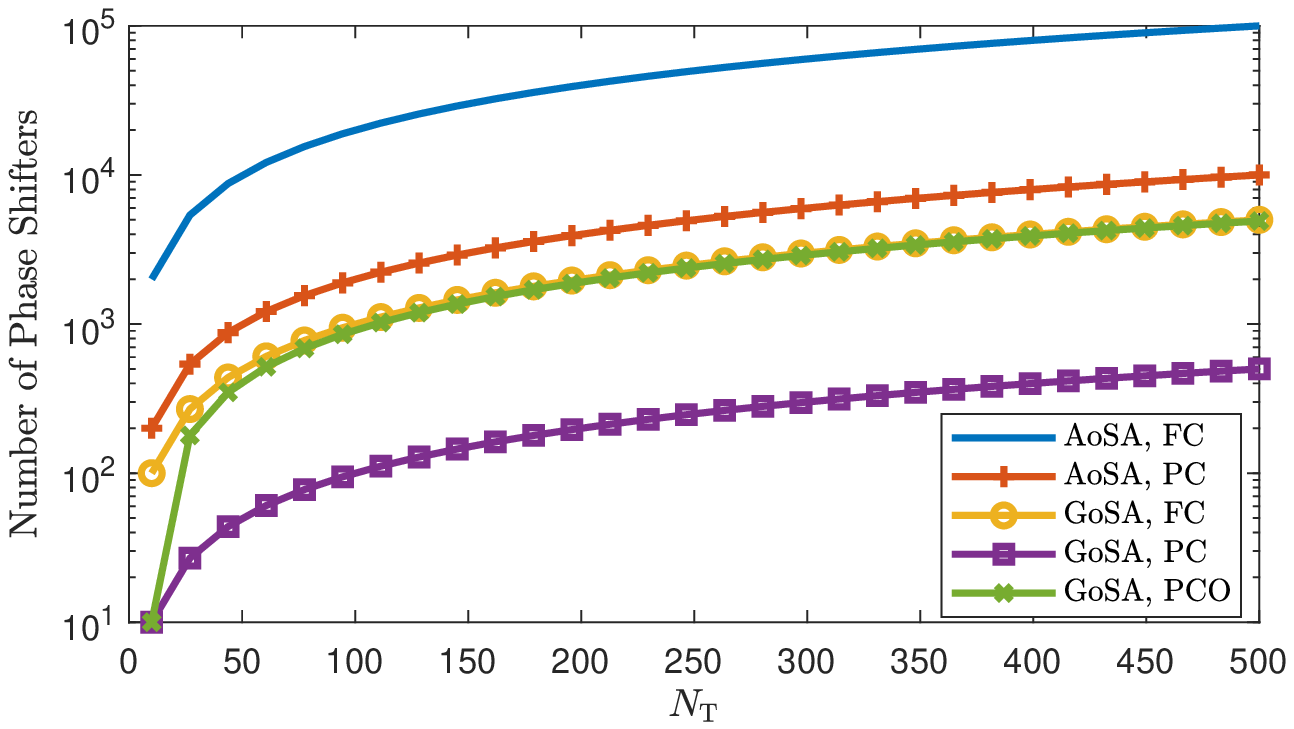} 	\label{fig_PhaseShiftNt}} \\
		\subfloat[] {\includegraphics[draft=false,width=\columnwidth]{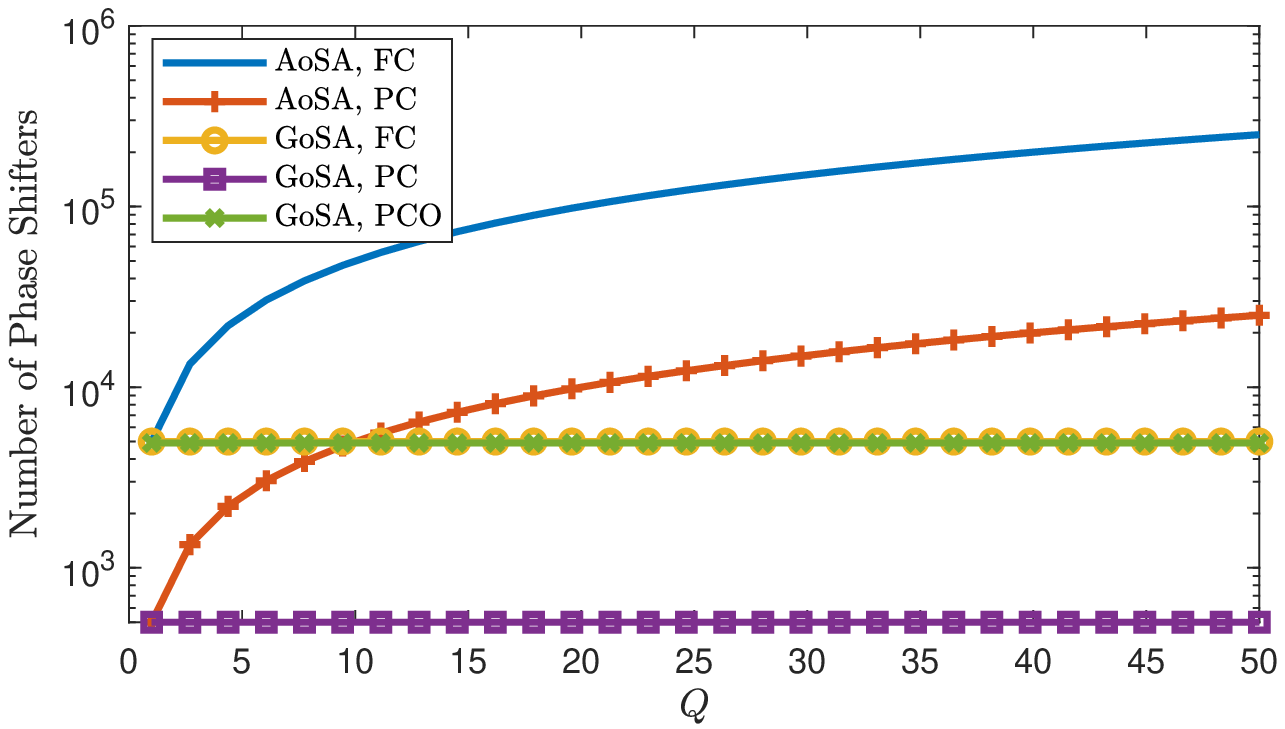} 	\label{fig_PhaseShiftQ}} 
		\caption{Number of phase-shifters versus (a) $N_\mathrm{T}$ when $Q=20$ and (b) $Q$ when $N_\mathrm{T} = 500$ for $N_\mathrm{RF} = 10$.  
		}
		\label{fig_NumPhaseShift}
	\end{figure}

	\section{Numerical Experiments}
	\label{sec:numexp}
	In this section, we evaluate the performance of the proposed hybrid beamforming approach for different array structures. The communications performance of the algorithms is evaluated in terms of spectral efficiency while the radar performance is presented with the beampattern analysis of the hybrid beamformers. Furthermore, we analyze the trade-off between both tasks by sweeping $\eta$ for $[0,1]$.  The hybrid beamformers are designed for fully-connected, partially-connected and PCO array structures. The proposed MMO-based approach is used to design PCO array. Then, it is compared with the partially-connected and fully-connected arrays, which employ the MO-based alternating minimization (MO-AltMin)~\cite{hybridBFAltMin} and triple AltMin (TAltMin) approach in~\cite{radarCommLiuICASSP2019}, respectively, while the fully digital unconstrained beamformers are used as a benchmark~\cite{heath2016overview}.
	
	\begin{figure}[t]
		\centering
		{\includegraphics[draft=false,width=\columnwidth]{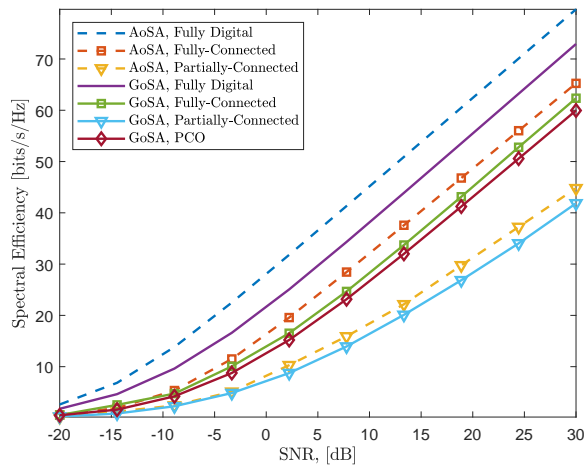} } 
		\caption{\color{black}Spectral efficiency versus SNR for CSI-based beamforming when $\eta = 0.5$. 
		}
		\label{fig_SE_SNR_ins}
	\end{figure}

	In the simulations, unless stated otherwise, we select the operating frequency as $f_c = 300$ GHz \textcolor{black}{with $M=64$ and $B=15$ GHz bandwidth}, which is in low-THz band ($100$ GHz - $1$ THz) and applicable for long range radar (LRR) ($\sim 150$ m)~\cite{gashinovaLowTHz1}. We also select $\Delta =\Delta_\mathrm{x} = \Delta_\mathrm{y} = \lambda/2$ and $\delta =\delta_\mathrm{x} = \delta_\mathrm{y} = \lambda/4$, {\color{black}where $\lambda$ denotes the wavelength corresponding to the carrier frequency}. At the TX and RX, $N_{\mathrm{T}_\mathrm{x}} = N_{\mathrm{T}_\mathrm{y}} = 32$ ($N_\mathrm{T} = 1024$) and $N_{\mathrm{R}_\mathrm{x}} = N_{\mathrm{R}_\mathrm{y}} = 9$ ($N_\mathrm{R} = 81$) subarrays are used, respectively, with  $Q_\mathrm{x} =Q_\mathrm{y}= 3$ ($Q=9$). Thus, the resultant architecture forms a $729\times 9216$ ultra-massive MIMO transceiver. We assume that $N_\mathrm{RF} = 16$ RF chains are used at the TX to transmit $N_\mathrm{S}=4$ data streams to the RX via the THz channel which is assumed to include {\color{black}one LoS and four NLoS (i.e., $L=5$)} paths, where $\phi_l,\varphi_l \in [-150^\circ,150^\circ]$ and $\theta_l,\vartheta_l \in [70^\circ,90^\circ]$. The TX simultaneously generates beams towards both RX and $K=3$ radar targets located at $\{(60^\circ,70^\circ),(110^\circ,75^\circ),(140^\circ,80^\circ)\}$. For model-free approach, we consider 1-D scenario, i.e., the elevation angles of the targets are $90^\circ$ for simplicity.
	
	{\color{black}The learning model \texttt{DeepBF} is realized as a CNN with $11$ layers. The first layer is the input layer of size $N_\mathrm{T}\times (N_\mathrm{R}+K)\times 2$. The second, fourth and sixth layers are convolutional layers with $256$@$3\times 3$ filters. After first two convolutional layers, there is a max-pooling to reduce dimension by $2$. The seventh and ninth layers are fully connected layers with $1024$ units. The eighth and tenth layers are dropout layers with $50\%$ rate. Finally, the last layer is a regression layer of size $2N_\mathrm{T}N_\mathrm{S}$. Let $\mathcal{D}_i = (\boldsymbol{\Xi}_i, \boldsymbol{\xi}_i)$ be the $i$th input-output tuple of the training dataset for $i = 1,\dots, \textsf{D}$, where $\textsf{D}= |\mathcal{D}|$ denotes the number of samples in the dataset. In order to generate the training dataset, we consider $Z_\mathrm{C}=10^2$ channel realizations with the aforementioned channel statistics and $K=3$ radar target directions, which are generated uniform randomly from the interval $[-50^\circ,50^\circ]$ with $1^\circ$ resolution for $Z_\mathrm{R} = 10^4$ realizations. Once the input data $\boldsymbol{\Xi}_i$ is prepared as described in Section~\ref{sec:modelFree}, the optimization problem in (\ref{problem2OFDM}) is solved for each input data, then the corresponding output label, i.e., $\boldsymbol{\xi}_i$ is computed for $i = 1,\dots, \textsf{D}$ in an offline manner. As a result, the resulting  dataset is comprised of $\textsf{D} = Z_\mathrm{C}Z_\mathrm{R} = 10^6$ samples of size $N_\mathrm{T}\times (N_\mathrm{R}+K)\times 2$. The cost function for \texttt{DeepBF} is the MSE between $\mathcal{L}(\boldsymbol{\Xi}_i, \boldsymbol{\theta})$ and $\boldsymbol{\xi}_i$. The \texttt{DeepMUSIC} model is constructed as described in~\cite{elbir_DL_MUSIC}.  Then, the learning models are realized in MATLAB on a PC with $2304$ GPU cores. We use the stochastic gradient descent (SGD) algorithm with momentum of $0.9$  and  update the network parameters with learning rate $0.001$ when the mini-batch size is $64$. }


	Fig.~\ref{fig_NumPhaseShift} shows the number of phase-shifters with respect to $N_\mathrm{T}$ and $Q$ for different array structures, i.e., AoSA and GoSA, respectively. The fully-connected structures employ $N_\mathrm{T}QN_\mathrm{RF}$ and $N_\mathrm{T}N_\mathrm{RF}$ phase-shifters for AoSA and GoSA, respectively, while  the partially-connected structures are more efficient since only $N_\mathrm{T}Q$ and $N_\mathrm{T}$ phase-shifters are used for AoSA and GoSA. Compared to AoSA, the proposed GoSA structure employs much less phase-shifters than that of AoSA for $Q \geq N_\mathrm{RF}$ and they become equal if $Q =1$. Thus, GoSA is much more energy-efficient than AoSA.  While GoSA provides lower hardware complexity, it has  slightly poorer spectral efficiency performance, which is ameliorated via the PCO structure by increasing the number of phase-shifters from $N_\mathrm{T}$ (non-overlapped) up to $N_\mathrm{RF}(N_\mathrm{T} - N_\mathrm{RF}+1)$ (fully-overlapped). Nevertheless, the fully-overlapped or fully-connected GoSAs still have lower phase-shifters than that of AoSA with partially-connected structure.

	\begin{figure}[t]
		\centering
		{\includegraphics[draft=false,width=\columnwidth]{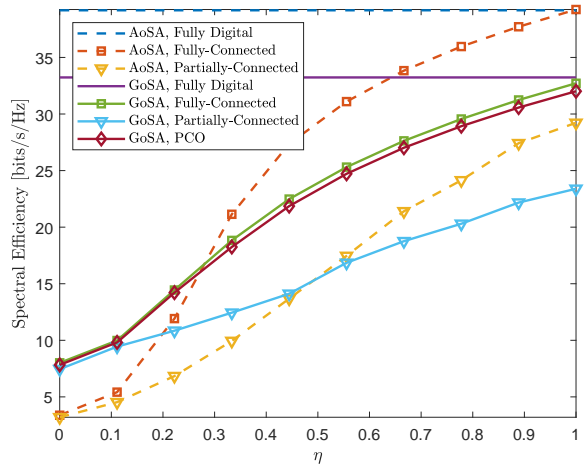} } 
		\caption{\color{black}Spectral efficiency versus $\eta$ for SNR$=10$ dB. 
		}
		\label{fig_SE_eta}
	\end{figure}

	%

	\begin{figure}[t]
		\centering
		\subfloat[]{\includegraphics[draft=false,width=\columnwidth]{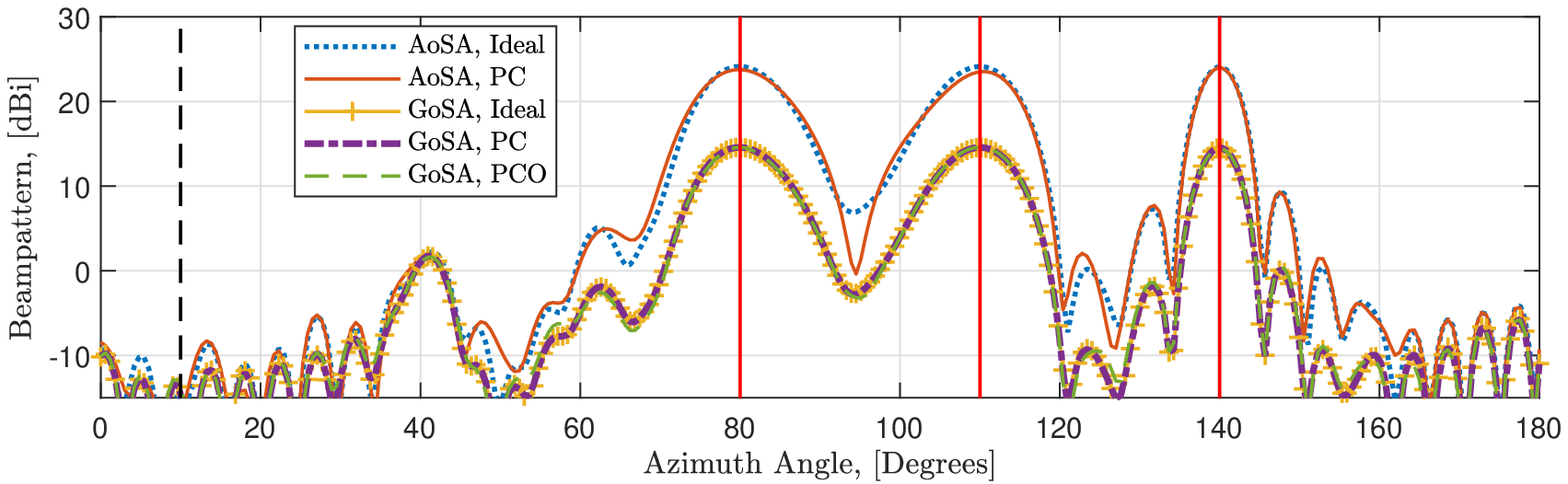} } \\
		\subfloat[]{\includegraphics[draft=false,width=\columnwidth]{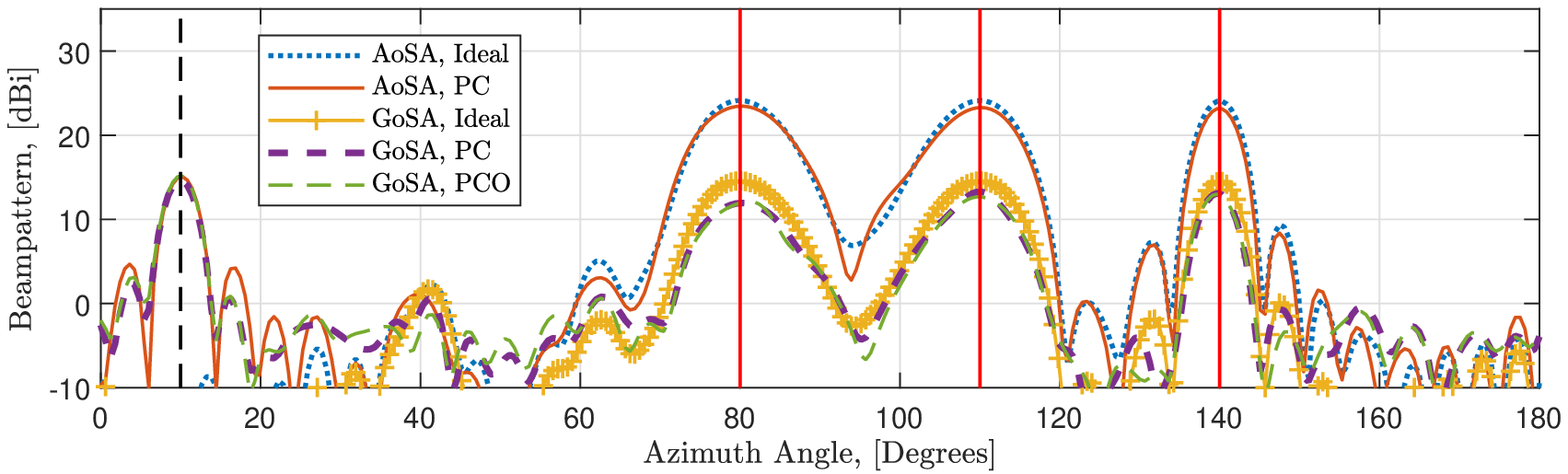} } \\
		\subfloat[]{\includegraphics[draft=false,width=\columnwidth]{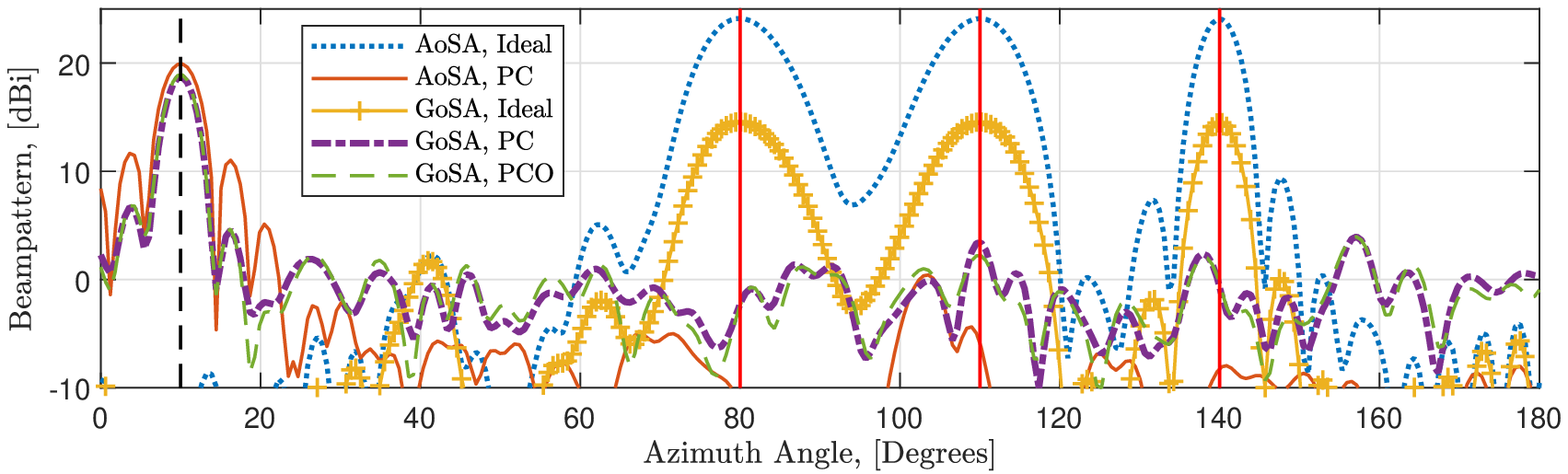} } 
		\caption{\color{black}Beampattern in the azimuth plane for (a) $\eta = 0$, (b) $\eta = 0.5$ and (c) $\eta = 1$, when the radar targets are located at $\{(80^\circ,90^\circ),(110^\circ,90^\circ),(140^\circ,90^\circ)\}$ and the LoS communications paths is received from $(10^\circ,90^\circ)$, which are depicted by solid and dashed vertical lines, respectively. 
		}
		\label{fig_BP_az}
	\end{figure}

	\begin{figure}[t]
		\centering
		{\includegraphics[draft=false,width=\columnwidth]{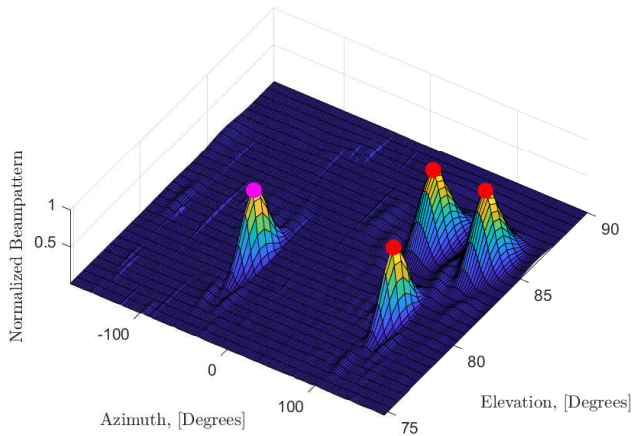} }  
		\caption{\color{black} Beampattern in 2-D plane for $\eta = 0.5$. The radar targets are located at $\{(80^\circ,85^\circ),(110^\circ,80^\circ),(140^\circ,85^\circ)\}$ and the LoS communication path is received from $(-50^\circ,80^\circ)$, which are illustrated with red and pink colors, respectively. 
		}
		\label{fig_BP_az_el}
	\end{figure}
	
	
	Fig.~\ref{fig_SE_SNR_ins} shows the spectral efficiency with respect to SNR for CSI-based hybrid beamforming when $\eta = 0.5$. We observe that  GoSA performs slightly lower than AoSA structure while using $Q=9$ times less phase-shifters, which significantly lowers the hardware complexity of ultra-massive MIMO system. While partially-connected structures have the lowest hardware complexities, they perform the worst as compared to the fully-connected case. The GoSA with PCO improves the spectral efficiency by employing relatively more phase-shifters which still less than that of AoSA. The gap between the unconstrained (fully digital) and hybrid beamformers is large due to the trade-off between radar and communications tasks with $\eta = 0.5$. 
	
	In Fig.~\ref{fig_SE_eta}, the spectral efficiency is presented with respect to $\eta$, wherein we note that as $\eta \rightarrow 1$, the spectral efficiency for the fully-connected, partially-connected and PCO approaches to the performance of unconstrained beamformer, i.e., $\mathbf{F}_\mathrm{C}[m]$. When $\eta \rightarrow 0$, then the RF precoder $\mathbf{F}_\mathrm{RF}$ generates the beams towards the radar targets only, thus the spectral efficiency is reduced. As a result, the selection of $\eta$ is critical. In practice, $\eta$ is increased if the communications task is more critical than tracking the targets or when there is no target. Conversely, lower $\eta$ is selected if the radar task demands more resources, e.g., more transmit power is required depending on the range of the radar targets.

	\begin{figure}[t]
		\centering
		{\includegraphics[draft=false,width=\columnwidth]{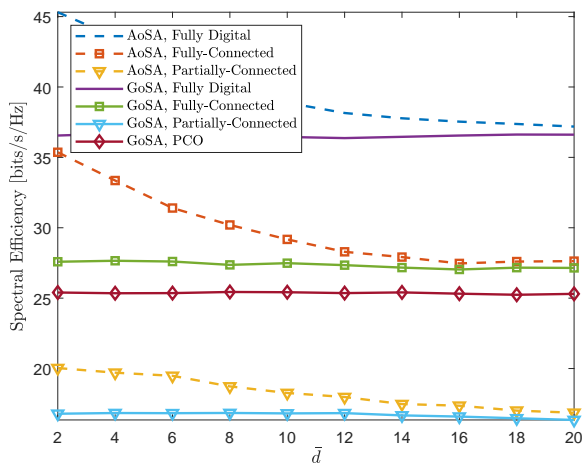} } 
		\caption{\color{black}Spectral efficiency versus $\bar{d}$ for SNR$=10$ dB when $\eta = 0.5$. 
		}
		\label{fig_SE_SNR_dbar}
	\end{figure}

	\begin{figure}[t]
		\centering
		{\includegraphics[draft=false,width=\columnwidth]{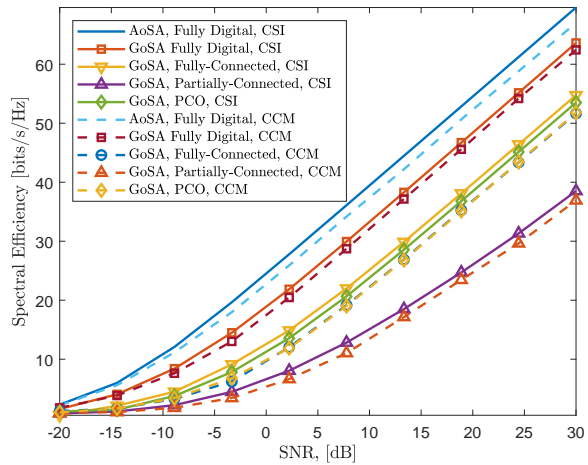} } 
		\caption{\color{black}Spectral efficiency versus SNR for channel covariance matrix-based beamforming when $\eta = 0.5$. 
		}
		\label{fig_SE_SNR_CCM}
	\end{figure}

	We illustrate the beampattern of the designed hybrid beamformers in Fig.~\ref{fig_BP_az} and Fig.~\ref{fig_BP_az_el} for 1-D and 2-D angle distributions, respectively. In Fig.~\ref{fig_BP_az}, the beampatterns are presented for $\eta = 0$, $\eta = 0.5$ and $\eta = 1$, where we assume that the all of the paths have the elevation angle of $90^\circ$. The ideal beampatterns correspond to the radar-only beamformer $\mathbf{F}_\mathrm{R}$ for AoSA and GoSA structures. We note that for $\eta =0$ ($\eta = 1$) all the beams are generated towards the radar targets (the RX), respectively, while $\mathbf{F}_\mathrm{RF}$ generates $K+1$ beams towards both targets and the RX when $\eta =0.5$. The proposed GoSA PCO structure provides lower side lobes and narrower beams towards both RX and radar targets as compared to the other algorithms. The 2-D angular distribution is illustrated in Fig.~\ref{fig_BP_az_el} for $\eta =0.5$, where we present the beampattern corresponding to the GoSA with PCO. We observe that the proposed hybrid beamforming approach accurately generates beams towards both targets and RX paths in 2-D angular space. This is provided with the 2-D structure of the antenna array in two dimensions.
	
	While the design of the antenna array is straightforward in AoSA case by selecting the antenna spacing as $\lambda/2$, the selection of $\delta$ is critical for the GoSA structure illustrated in Fig.~\ref{figHB_TX}. In Fig.~\ref{fig_SE_SNR_dbar}, we present the spectral efficiency performance with respect to $\bar{d}$, where $\bar{d} = \frac{\lambda}{\delta}$. As $\bar{d}$ increases, we reduce the antenna element spacing in the subarrays of GoSA. Specifically, when $\bar{d} = \infty$, we have $\delta = 0$, thus the $Q_\mathrm{x}\times Q_\mathrm{y}$ antennas in each subarray become co-located. We infer this from Fig.~\ref{fig_SE_SNR_dbar} as the performance of AoSA approaches to that of GoSA as $\bar{d}$ increases. While slight performance loss is observed from AoSA with partially-connected array, the performance of the fully-connected structure and the fully digital beamformers significantly reduce as $\bar{d}$ increases. As a result, this figure is helpful when designing the GoSA because of the improvement in the spectral efficiency by changing $\bar{d}$ while the lower limit is $\bar{d}=2$ (i.e., $\delta = \lambda/2$) to avoid spatial aliasing among the antennas.

	\begin{figure}[t]
		\centering
		{\includegraphics[draft=false,width=\columnwidth]{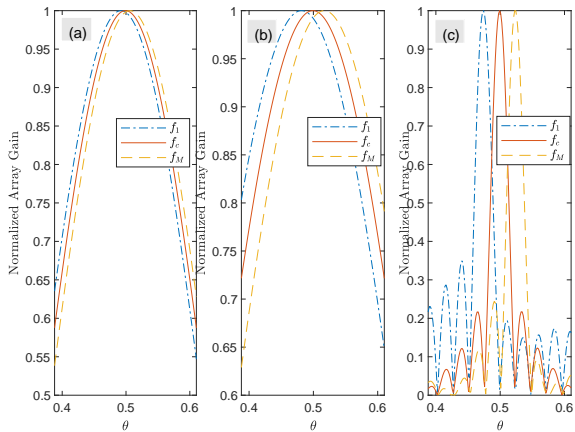} } 
		\caption{\color{black}Normalized array gain with respect to physical direction $\theta$ at lowest, center and highest subcarrier frequencies (i.e., $f_1$, $f_c$ and $f_M$) for (a) $f_c = 3.5$ GHz, $B = 0.1$ GHz; (b) $f_c = 28$ GHz, $B = 2$ GHz; and (c) $f_c=300$ GHz, $B = 30$ GHz when $M = 128$. 
		}
		\label{fig_BeamSplit}
	\end{figure}

	\begin{figure}[t]
		\centering
		{\includegraphics[draft=false,width=\columnwidth]{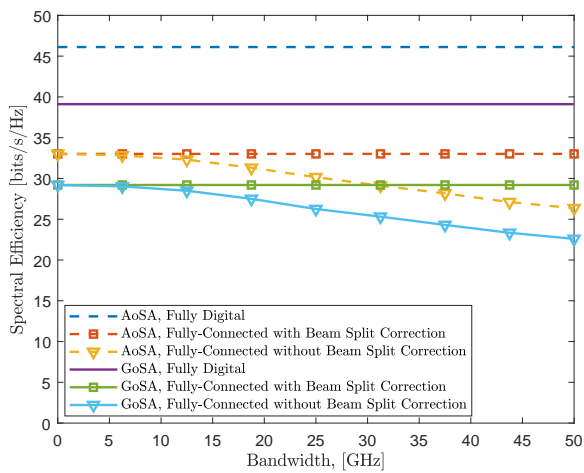} } 
		\caption{\color{black}Spectral efficiency versus bandwidth when $\eta = 0.5$. 
		}
		\label{fig_SE_SNR_BeamSplit}
	\end{figure}

	Fig.~\ref{fig_SE_SNR_CCM} shows the spectral efficiency of the competing algorithms for both CSI- and channel covariance matrix-based beamforming when $\eta = 0.5$. A slight performance loss is observed for all channel covariance matrix-based approaches due to loss of precision in the angle and path gain information while less channel overhead is involved in channel covariance matrix-based beamforming. In particular, in the CSI-based approach, the RX should feedback $N_\mathrm{R}\times N_\mathrm{T}$ ($81\times1024 $) channel matrix whereas only the angle and path information needs to be sent to the TX in the channel covariance matrix so that $\mathbf{C}[m]$ is constructed as in (\ref{covarianceApp}).

	{\color{black}Fig.~\ref{fig_BeamSplit} shows the effect of frequency on the generated beams for (a) sub-6 GHz, (b) mmWave and (c) THz MIMO systems. We observe that different beams point to very close physical directions at low frequencies while the beam split occurs at THz, wherein the main lobes corresponding to the lowest/highest and center subcarrier frequencies do not overlap.} We also present the effect of beam split on spectral efficiency with respect to bandwidth in Fig.~\ref{fig_SE_SNR_BeamSplit}. A severe loss in the spectral efficiency is observed when the bandwidth is large (i.e., $>20$ GHz for $f_c = 300$ GHz). This arises from the use of frequency-independent analog beamformer, which causes a misalignment of generated the beams at different subcarriers, hence degrades the spectral efficiency. This loss can be effectively mitigated by the proposed beam split correction technique, which tunes the phase mismatches in the analog beamformer due to the use of a single frequency, i.e., $f_c$.
	
	\begin{figure}[t]
		\centering
		{\includegraphics[draft=false,width=\columnwidth]{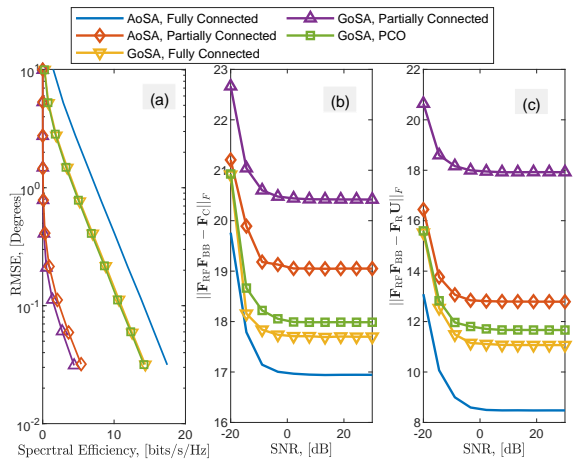} } 
		\caption{(a) Radar target direction RMSE versus spectral efficiency, and the cost function  for (b) communications- and (c) radar- only beamformer.}
		\label{fig_JRC}
	\end{figure}
	
	{\color{black}Fig.~\ref{fig_JRC}a-c shows the performance of the hybrid beamformers in terms of radar target direction estimation together with the cost function for communications- and radar- only beamformers, respectively. In Fig.~\ref{fig_JRC}a, SNR is swept for $[-20,30]$ dB and the corresponding direction root-mean-squared error (RMSE) and the spectral efficiency are computed.  It can be seen that both radar (direction RMSE) and communication (spectral efficiency) performance improves proportionally as SNR increases while the partially connected arrays perform poorer than the fully connected ones for both AoSA and GoSA structures. Nevertheless, the proposed GoSA PCO array exhibits satisfactory performance for both radar and communications. }

\begin{figure}[t]
		\centering
		{\includegraphics[draft=false,width=\columnwidth]{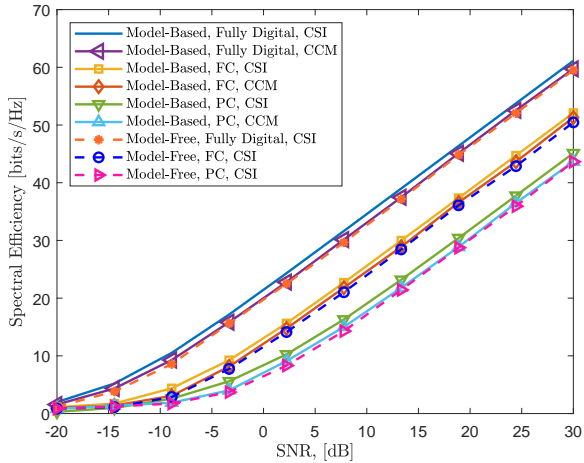} } 
		\caption{Spectral efficiency for model-based and model-free hybrid beamformer design when $\eta = 0.5$. 
		}
		\label{fig_SE_SNR_ModelFree}
	\end{figure}

	Fig.~\ref{fig_SE_SNR_ModelFree} shows the spectral efficiency comparison of model-based and model-free techniques when GoSA structure is used. The simulations are averaged over,  $500$ Monte Carlo trials, each of which is conducted for different realization of radar target angles. {\color{black}Note that we considered narrowband scenario in this simulation due to the memory limitations of the computation platform used for model training, while the results can be generalized for wideband scenario as presented in~\cite{elbir2019online}.} While a slight loss is observed for the model-free techniques compared to CSI-based beamforming, they have close performance to the channel covariance matrix-based methods. Another advantage of the model-{\color{black}free} approach is computational complexity thanks to its implementation via parallel processing units, such as GPUs. {\color{black}The performance of the learning model depends on the size of the dataset, which should cover a large portion of the whole input space. In case of smaller datasets, transfer learning-based approaches may be used~\cite{elbir2020TL}.} Based on simulations for the aforementioned TX-RX settings, the computation time for MO, TAltMin, \texttt{DeepMUSIC} and \texttt{DeepBF} are $2.124$, $0.68$, $0.0036$ and $0.0058$ seconds, which shows the advantage of model-free techniques.

	\section{Summary}
	\label{sec:summ}
	In this paper, we introduced a THz ultra-massive MIMO JRC architecture and investigated model-based and model-free hybrid beamforming techniques. To lower the hardware complexity critical in THz systems, we proposed GoSA ultra-massive MIMO architecture. 	We developed hybrid beamforming via PCO structures to provide a trade-off between higher spectral efficiency and hardware complexity in terms of the number of phase-shifters. The hybrid beamformers for THz JRC system are designed relying on both CSI and channel covariance matrix of the wireless channel information between the TX and the RX. The computation times for beamformer design could be prohibitively high for ultra-massive MIMO THz systems. We addressed this by suggesting a model-free DL-based approach. 
	
	We evaluated the performance of the proposed methods in terms of spectral efficiency and radar beampattern. We demonstrated that GoSA provides less hardware complexity compared to full array and AoSA structures. \textcolor{black}{To mitigate the beam split effect, we also introduce hardware-efficient approach by correcting the phases of the frequency-independent beamformers.} Compared to CSI-based beamforming, channel covariance matrix-based approach has a slight performance loss, while the latter enjoys less channel overhead. The model-free method is advantageous in terms of computational complexity and exhibits approximately $500$ times lower computation time as compared to the MO-based approaches, while maintaining  spectral efficiency performance close to that of channel covariance matrix-based technique.

	\appendices

	{\color{black}
		\section{Beam Split Correction}
		\label{appBeamsplit}
		In mmWave wideband hybrid beamforming, the analog beamformers are usually designed with respect to a single frequency, (i.e., $f_c$) while the baseband beamformers are frequency-dependent. Hence, the analog beamformers may point in different directions at different subcarriers because of ultra-wide bandwidth and large number of antennas. In particular, let $\mathbf{f}(\mathbf{\Omega}_l)$ be the beamforming vector corresponding to the $3\times 1$ spatial direction vector $\mathbf{\Omega}_l$  as defined in (\ref{svector}), then the beam generated by the frequency-independent beamformer is aligned with the frequency-dependent physical direction vector $\widetilde{\mathbf{\Omega}}_{l}$ as
		\begin{align}
		\widetilde{\mathbf{\Omega}}_{l} = {\mathbf{\Omega}}_{l} \Delta_m,
		\end{align}
		where $\Delta_m = \frac{f_c}{f_m}$ denotes the relative frequency compared with the central frequency $f_c$~\cite{thz_beamSplit}. To eliminate the effect of beam splitting, the phases of the frequency-independent beamformer should be corrected.  Let $\hat{\mathbf{F}}_\mathrm{RF}(\mathbf{\Omega})$  be the solution of hybrid beamforming problem in (\ref{problem2OFDM}) and $\mathbf{\Omega}$ represents the spatial directions in the analog beamformers. Then, the frequency-dependent analog beamformer is
		\begin{align}
		\label{beamSplit1}
		\hat{\mathbf{F}}_\mathrm{RF}^\mathrm{c}[m] = \hat{\mathbf{F}}_\mathrm{RF}(\mathbf{\Omega} \Delta_m),
		\end{align}
		where $\hat{\mathbf{F}}_\mathrm{RF}^\mathrm{c}[m]$ points to $\mathbf{\Omega}$ for all $m\in\mathcal{M}$. The beam split correction operation in (\ref{beamSplit1}) by simply multiplying the phase values of $\hat{\mathbf{F}}_\mathrm{RF}(\mathbf{\Omega})$ by $\Delta_m$. However, the implementation of $\hat{\mathbf{F}}_\mathrm{RF}^\mathrm{c}[m]$ is not efficient since it requires $M$ phase shifter network of size $N_\mathrm{T}N_\mathrm{RF}$. To mitigate this, the effect of $\Delta_m$ can be conveyed to the frequency-dependent baseband beamformers. The modified baseband beamformer at subcarrier $m$ becomes
		\begin{align}
		\label{beamSplitEquation}
		\hat{\mathbf{F}}_\mathrm{BB}^\mathrm{c}[m] = \big(\hat{\mathbf{F}}_\mathrm{RF}(\mathbf{\Omega})\big)^\dagger  \hat{\mathbf{F}}_\mathrm{RF}^\mathrm{c}[m] \hat{\mathbf{F}}_\mathrm{BB}[m].
		\end{align}
		Finally,  the beam split corrected hybrid beamformer can be realized as $\hat{\mathbf{F}}_\mathrm{RF}(\mathbf{\Omega})\hat{\mathbf{F}}_\mathrm{BB}^\mathrm{c}[m]$.
		
	}
	
	\color{black}
	\section*{Acknowledgements}
	The authors are sincerely grateful to the guest editor Prof. Robert W. Heath, Jr. and three anonymous reviewers whose valuable comments greatly helped in improving the manuscript.
	
	\color{black}
	\balance
	\bibliographystyle{IEEEtran}
	\bibliography{IEEEabrv,references_089}
	
\end{document}